%% file: main_aies.tex
\documentclass[draft]{article} 
\usepackage{aaai25}
\usepackage{times} 
\usepackage{helvet}
\usepackage{courier} 
\usepackage{xcolor}
\usepackage[hyphens]{url} 

\usepackage[pdftex,final]{graphicx}

\usepackage{array}
\usepackage{float}
\usepackage{rotating}
\usepackage{enumitem}

\usepackage{natbib} 
\usepackage{caption} 
\usepackage{booktabs}
\usepackage{amssymb}
\usepackage{algorithm}
\usepackage{algorithmic}

\usepackage{makecell}
\usepackage{subcaption}

\usepackage[T1]{fontenc}

\usepackage{newfloat}
\usepackage{listings}
\DeclareCaptionStyle{ruled}{labelfont=normalfont,labelsep=colon,strut=off} 
\floatstyle{ruled}
\newfloat{listing}{tb}{lst}{}
\floatname{listing}{Listing}

\newenvironment{usecase}[6]{%
  \vspace{0.5em}%
  \begingroup
  {\large\bfseries #1.\ #2}\\[0.25em]
  \begin{center}
    \vspace{-0.25em}
    \includegraphics[width=0.75\linewidth]{#5}\\
    \captionof{figure}{#2\ diagram.}\label{#6}
    \vspace{-0.5em}
  \end{center}
  \noindent\textbf{Sector:} #3\\
  \noindent\textbf{Domain:} #4\\[0.25em]
}{%
  \endgroup
  \vspace{0.5em}
}

\title{Documenting Deployment with Fabric: A Repository of Real-World AI Governance}
\author {
    Mackenzie Jorgensen\equalcontrib \textsuperscript{\rm 1},
    Kendall Brogle\equalcontrib \textsuperscript{\rm 1},
    Katherine M. Collins\textsuperscript{\rm 2},
    Lujain Ibrahim\textsuperscript{\rm 3},
    Arina Shah\textsuperscript{\rm 4},
    Petra Ivanovic\textsuperscript{\rm 5},
    Noah Broestl\textsuperscript{\rm 2,6},
    Gabriel Piles\textsuperscript{\rm 7}, 
    Paul Dongha\textsuperscript{\rm 8},
    Hatim Abdulhussein\textsuperscript{\rm 9,10}, 
    Adrian Weller\textsuperscript{\rm 1,2},
    Jillian Powers\textsuperscript{\rm 11},
    Umang Bhatt\textsuperscript{\rm 1,2,4}\footnote{Correspondence to: \{ncb49,usb20\}@cam.ac.uk}
}
\affiliations {
    \textsuperscript{\rm 1}The Alan Turing Institute,
    \textsuperscript{\rm 2}University of Cambridge, 
    \textsuperscript{\rm 3}University of Oxford, 
   \textsuperscript{\rm 4}New York University, 
    \textsuperscript{\rm 5}Trustwise, 
    \textsuperscript{\rm 6}Boston Consulting Group, 
    \textsuperscript{\rm 7}HURIDOCS,  
    \textsuperscript{\rm 8}NatWest, 
    \textsuperscript{\rm 9}University of Surrey, 
    \textsuperscript{\rm 10}Health Innovation Kent Surrey Sussex, 
 \textsuperscript{\rm 11}Slalom
}

\begin{document}

\maketitle

\begin{abstract}
Artificial intelligence (AI) is increasingly integrated into society, from financial services and traffic management to creative writing. Academic literature on the deployment of AI has mostly focused on the risks and harms that result from the use of AI. We introduce Fabric, a publicly available repository of deployed AI use cases to outline their governance mechanisms. Through semi-structured interviews with practitioners, we collect an initial set of 20 AI use cases. In addition, we co-design diagrams of the AI workflow with the practitioners. We discuss the oversight mechanisms and guardrails used in practice to safeguard AI use. The Fabric repository includes visual diagrams of AI use cases and descriptions of the deployed systems. Using the repository, we surface gaps in governance and find common patterns in human oversight of deployed AI systems. We intend for Fabric to serve as an extendable, evolving tool for researchers to study the effectiveness of AI governance. 
\end{abstract}

\section{Introduction}
Artificial intelligence (AI) is now deployed at scale, stretching from wayfinding through cities to determining creditworthiness to recommending music~\cite{gartner2019aiadoption,faheem2021ai, grandview2024aimarket,mokoena2025analysis}. 
Much academic literature to date has rightly focused on either advancing system capabilities or documenting risks and failures associated with system deployments. Multiple public databases, like the Responsible AI Collective's AI Incident Database~\cite{mcgregor2021preventing}, the OECD's AI Incidents and Hazards Monitor~\cite{oecd2025aiincidents}, and MIT’s AI Risk Repository~\cite{slattery2024ai}, highlight harms of AI system failures.  

While these repositories capture AI risks, there is a dearth of knowledge on how AI systems are deployed and what governance mechanisms make such deployment safe and possible \cite{aigovernancelibrary}. To address which governance mechanisms make AI deployment less risky, the United States' National Institute of Standards and Technology (NIST) released their AI Risk Management Framework~\cite{ai2024artificial}. Complementary efforts include the OECD's catalog of tools for trustworthy AI \cite{oecdtools}  and the UK government's portfolio of AI assurance techniques\cite{govukaiassurance}. However, the efficacy of these guidelines is unclear so far in practice.

AI systems are often quietly integrated into workforce technologies, public services, and consumer products without a shared understanding of how and when they work~\cite{john2021architecting,meng2025data}. In this paper, we begin to fill gaps in understanding how AI systems are used in practice and what governance surrounds their use. More specifically, we ask: how is governance imposed on deployed AI systems? To address this question, we release Fabric, a public repository of AI system deployments. The deployed use cases are categorized by their sector, oversight patterns, and other meta-data. 

To gather AI use cases, we conduct semi-structured interviews with AI practitioners, where we understand their AI workflows and how governance is applied. Crucially, we co-design a diagram of the underlying AI workflow with each practitioner. We present 20 AI use cases in our repository. For each use case, we describe the AI system, discuss the governance around its use, and provide the co-designed diagram. We analyze the repository to find commonalities across use cases and devise governance patterns, which we hope can guide future deployment of AI. We cast a wide net in our definition of ``AI systems,'' ranging from large language models (LLMs) to shallow neural networks to rule-based expert systems. 
We focus on tactics to operationalize AI governance, the control and oversight of deployed AI systems. 
Instead of theorizing which principles dictate governance at scale, we capture instances of AI governance from real-world use cases.

We aim to offer academics, industry professionals, and policymakers a landscape view of how AI systems can be governed by releasing the governance patterns from our use cases. Our initial collection of them is only the beginning; we envision the repository will grow with broader community engagement. This paper proceeds as follows: we outline related work in Section \ref{sec:related}; we describe our methodology in Section \ref{sec:methods}; we provide an overview of the Fabric repository and define the governance patterns that arise across the Fabric use cases in Section \ref{sec:usecases}; we discuss the governance considerations from the use cases in Section \ref{sec:disc}; lastly, we conclude in Section \ref{sec:conc} with our hopes for future iterations of work.

\section{Related Work}
\label{sec:related}

AI is increasingly deployed across a wide range of decision-making contexts, from high stakes domains (e.g., healthcare~\cite{esteva2017dermatologist} and immigration~\cite{booth2024}) to lower-risk applications (e.g., personalized writing assistance~\cite{hwang2023ai})---even potentially forming the base of new kinds of collaborative cognition in human-AI thought partnerships~\citep{collins2024building}. While AI systems offer benefits in terms of efficiency and scalability, they also introduce harmful and discriminatory risks. Thus, we need ways to govern these AI systems to mitigate for those harms and risks.

\subsection{Proposals for AI Governance}
AI governance can be understood as the combination of institutional, technical, and procedural mechanisms that ensure the legal, ethical, and safe development and
deployment of AI systems~\cite{jobin2019global,schneider2023artificial}, which may impact not just individuals but collective networks of people~\citep{ brinkmann2023machine, collins2025revisiting}. In response to risks (e.g., from biased AI systems~\cite{mehrabi2021survey}), the literature proposes a range of governance mechanisms that we non-exhaustively outline.

Proposals span both legal-institutional structures and procedural mechanisms that shape how AI is used after deployment.
Legal frameworks, such as the European Union's (EU) AI Act and the EU General Data Protection Regulation (GDPR) mandate different forms of human oversight and transparency ~\cite{eu_gdpr_2016,edwards2021eu,euai2024,ai2024artificial}.
Within the United States, NIST has created an AI risk management framework that provides recommendations but is not legally enforceable \cite{dotan2024nist}.
Beyond legal structures, the literature outlines several normative governance mechanisms that can be expressed post hoc, especially when the organization has procured third party AI systems. 
These include human-in-the-loop review \cite{green2022flaws}, escalation protocols \cite{Upmann2024AIescalation}, logging requirements \cite{cheong2024transparency}, oversight by default \cite{CIPL2024accountableAI}, post-hoc monitoring \cite{mokander2023operationalising}, and workflow-based gate-keeping \cite{KayeDixonGellman2023}. 
Unfortunately, such mechanisms are often described abstractly with little guidance as to how they can be implemented in practice. 

Complementing these mechanisms, Shao et~al \cite{shao2025future} introduces a Human Agency Scale (HAS) that provides a shared, empirical language for calibrating the degree of human involvement with AI agents. Their framework reveals mismatches between worker preferences and expert assessments, mapping tasks into actionable zones of automation and augmentation. These insights can directly inform the design of oversight and escalation mechanisms so that governance aligns more closely with human preferences. 

\subsection{Addressing Gaps in the Literature}
While existing initiatives, such as the Responsible AI Collective's AI Incident Database, the OECD's AI Incidents and Hazards Monitor, and MIT's AI Risks Repository, document failures and risks associated with AI system deployment~\cite{mcgregor2021preventing,oecd2025aiincidents,slattery2025airiskrepositorycomprehensive}, these resources are inherently reactive. They flag where governance has failed but do not provide insight into how governance is structured in operational settings. Similarly, documentation methods, like model cards~\cite{mitchell2019model} and data sheets~\cite{holland2020dataset,gebru2021datasheets, frieder2024data} promote transparency but stop short of capturing how oversight, escalation, or accountability is embedded into AI workflows.

Normative frameworks propose governance mechanisms, such as selective abstention \cite{chow1970optimum, cortes2017adanet}, confidence disclosures \cite{shortliffe1975model,doshi2017towards, le2023explaining}, and transparency prompts \cite{wachter2017right,ananny2018seeing, kaminski2021right}: all of which are grounded in responsible AI principles. While these proposals offer valuable guidance, there is limited empirical research showing how such mechanisms are actually implemented for real-world AI systems, or how they are practically integrated into institutional oversight structures. As a result, a critical gap remains, examining how AI governance is enacted in real-world AI systems.

This paper addresses this gap by documenting governance mechanisms across a set of deployed AI systems. By co-designing AI workflow diagrams with practitioners, we visualize how oversight, institutional policies, and accountability checkpoints operate in real-world AI use. The Fabric repository offers glimpses into what governance looks like in practice, informing both researchers and practitioners on how to bridge the divide between principles and implementation.

\subsection{Repository Initiatives}
The need for repositories of AI governance practices is a widely recognized necessity by governments, researchers and industry. In fact, U.S federal initiatives, including the National AI Initiative Act, recent executive actions, and the Department of Defense's Task Force LIMA, have called for public registries of AI uses cases and governance practices to support responsible deployment \cite{DoDTaskForceLima2024, EO13960_2020, mckernon2024ai, NARA2025AIInventory, OMB2024AIInventory}. Globally, repositories, such as the OECD AI Policy Observatory~\cite{oecd_ai_observatory}, UNESCO's Ethics of AI platform~\cite{unesco_ai_observatory}, and Singapore's Model AI Governance Framework~\cite{singapore_ai_framework}, offer centralized collections of AI-related laws, frameworks, and case studies. Industry and academic efforts have followed suit, including IBM's AI Governance Fact-sheets~\cite{ibm_ai_factsheets} and the University of Technology Sydney's AI Governance Lighthouse Series \cite{uts_ai_governance_case}. 

While valuable, these repositories are typically limited to policy-level summaries, template-driven documentation, or high-level descriptive accounts. They often do not examine how governance can be operationalized in existing AI workflows via human oversight, fallback mechanisms, or institutional constraints. The AI Governance Library~\cite{aigovernancelibrary} complements these repositories by offering a practitioner focused, blog style, archive of tools intended to provide professionals with practical strategies for deploying AI systems. 
Fabric differentiates itself from these repositories and addresses an unmet need by documenting governance structures as they are enacted. Rather than focusing solely on compliance or documentation, Fabric captures the workflow of deployment through co-designed diagrams and structured summaries of oversight mechanisms. In doing so, Fabric builds upon previous initiatives while providing a unique workflow-level, empirical, and cross-sectoral view of AI governance in action.  

\begin{figure*}[h]
    \centering
\includegraphics[width=0.8\linewidth]{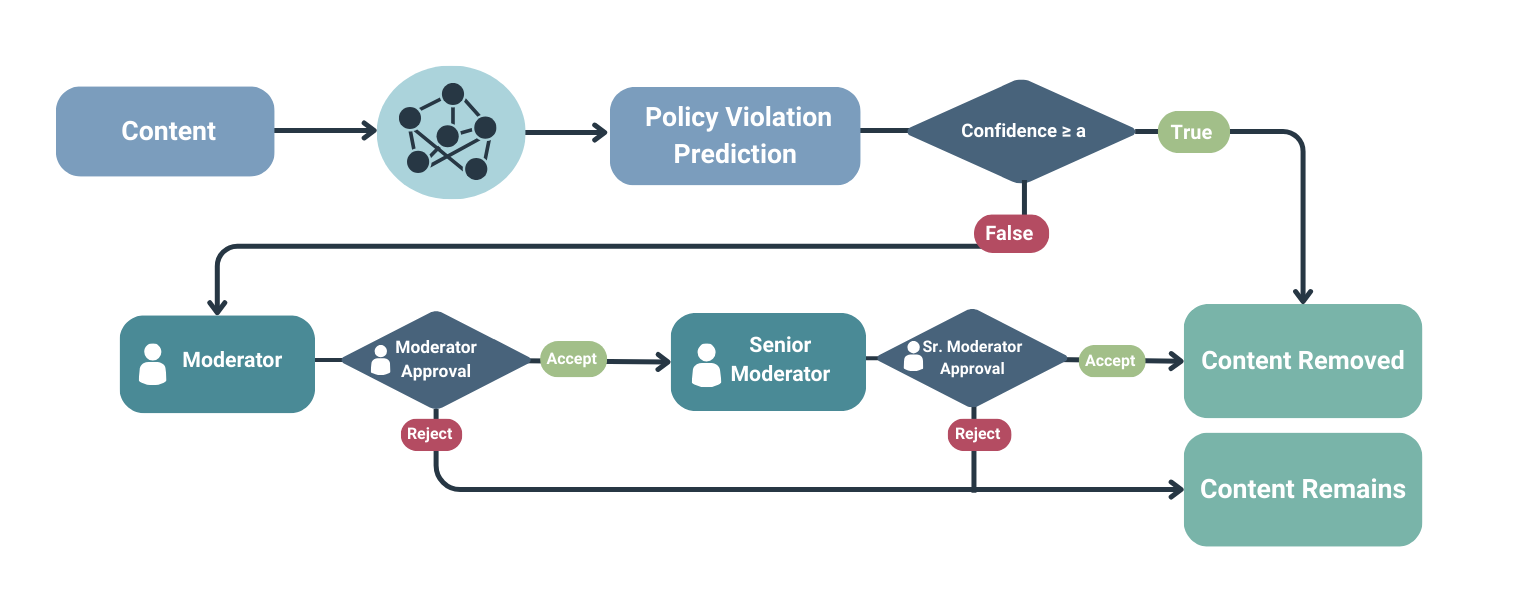}
    \caption{An example AI workflow for a hypothetical content moderation system deployed by a media platform to evaluate user content. }
    \label{fig:ex_case}
\end{figure*}

\section{Methodology} 
\label{sec:methods}
To address the lack of empirical understanding of how governance is implemented in practice, we set out to document real-world AI deployment with Fabric. We focus on how oversight is embedded within workflows and how AI systems interact with existing infrastructure and operations at organizations. To accomplish this, we use qualitative, semi-structured interviews~\cite{gubrium2002handbook} and collaborative co-design~\cite{poinet2020supporting} to depict the elements that shape system governance. Co-design, also dubbed prototyping, is a well-accepted methodology for creating artifacts alongside a participant~\cite{buchenau2000experience,madden2014probes,camburn2017design}: here, we work with an AI practitioner to create AI workflow diagrams similar to that shown in Figure~\ref{fig:ex_case}.
We obtained ethics approval for this study from The Alan Turing Institute's Institutional  Ethics Review Board.

To gather an initial set of use cases, over the course of five months, we circulated more than 80 invitations to participate, not including those sent through institutional mailing lists. We conducted virtual interviews with practitioners, individuals directly involved in the development, deployment, or maintenance of AI systems within an organization. Each semi-structured interview lasted around 45 minutes. We include our interview questions in Appendix A. During the interview, we collaborated with the practitioner to create an AI workflow diagram. The practitioners described their workflow, while we mapped out an initial diagram to review with the practitioner. Each interview had at least one interviewer present (most had two): one person asking questions and the other iterating on the diagram based on practitioner feedback. Real-time collaboration let us clarify doubts immediately and incorporate the practitioner's feedback iteratively to ensure the diagram represented the workflow well. After the interview, we refined the diagram and wrote a use case summary to share with the practitioner for approval. Once we collected multiple use cases, we sought governance patterns that emerge across workflows.

\subsection{Sampling and Interview Subjects}

Participants were recruited through personal networks, research partnerships, and Slack channels. We expanded our pool of interviewees through snowball sampling~\cite{biernacki1981snowball,noy2008sampling, parker2019snowball}. We adopted a broad definition of ``AI system'' to capture a wide audience. Before each interview, participants received a detailed study overview and signed a consent form that outlined our ethics considerations and privacy measures. 
Given our sample size, we used pseudoanonymization during data collection and storage to preserve anonymity. We note that interviews were optionally recorded and later deleted, after transcriptions of them were pseudoanonymized and saved. 

In this paper and in the initial Fabric repository, we do not disclose the names of organizations from which the use cases came, unless they specifically asked and provided us with consent to share this information. Each use case is categorized by organizational sector (e.g., private, public, and civil society) and domain, and all identifying information is stored securely and only accessible to the research team.

\subsection{Diagram Design}

We developed a ``diagram design'' approach to visually represent both technical workflows that include AI systems and the governance mechanisms at play. While we initially drew inspiration from software modeling frameworks, such as Unified Modeling Language (UML) diagrams \cite{jager1999using,jacobson2021unified}, we ultimately chose to use decision flowchart design to incorporate elements such as decision points, flow lines, and feedback loops \cite{yonyx2024explainingai,projectmanager2025decisionflowchart}. Decision flowchart design represents the interactive nature of AI systems and governance. 

As we interviewed practitioners, we updated the elements included in the diagrams because we needed to account for more complex workflows. We could not represent all elements of the workflows, so we prioritized certain common elements and included more fine-grained details in the use case summaries. We completed the diagram drafting on the Canva platform. A comprehensive key of the elements used in our diagrams is included in Figure 1 in Appendix B.  

To ground our methodology in a concrete example, we walk through a representative use case
for a content moderation system used by a hypothetical social media platform to identify content policy violations. 
In Figure~\ref{fig:ex_case}, the content is the input for an AI system. The AI system then evaluates if the content violates the platform's policy. In addition to a prediction of content violation, the AI system also outputs a confidence score for the prediction. The next step in the workflow is a decision point which occurs as a safety check on the system's confidence score.
A decision point is a place in the workflow that determines the next course of action. In this case, if the confidence score is above a predetermined threshold, then the content is automatically removed without any human intervention; thus, the AI system could act autonomously. However, if the score falls below the threshold, then the content is routed to a human moderator for review. If the moderator does not believe the content violates their policies so it is deemed safe, then the content remains on the platform. If the moderator believes the post violates their policy, then an additional moderator has to check this violating content before it is removed. The ``final output'' of this workflow is either the content is removed or the content remains, both of which is shown the diagram. We use these components to generate diagrams for the initial repository.

\subsection{Data Analysis}
To analyze what patterns of governance arose, our research team reviewed interview notes and AI workflow diagram summaries, and compared diagrams across use cases. As use case summaries and diagrams were built, we iteratively reviewed our findings to update the patterns of governance. In subsequent sections, we catalog the patterns found and report how many AI use cases fall under each pattern.

\section{Fabric: The Initial Inventory}
\label{sec:usecases}
The initial release of our Fabric repository is made up of 20 real-world use cases for AI, collected through semi-structured interviews with practitioners across a variety of sectors and types of organizations. Each use case centers around an AI system that is in active use within an organization. Repository entries include: the sector, domain, an AI workflow diagram that shows the operation AI is involved in, a brief description of the task, the intention behind deploying the system, its risks, patterns of governance, and an explanation of the AI workflow diagram.
Collectively, these use cases are intended to ground abstract theories on AI oversight in empirical reality, documenting how governance is enacted in practice.
In Appendix B, we show all 20 use cases from the initial Fabric repository, including their diagrams and descriptions. We also include summary statistics of the use cases in Table 1 in Appendix B.

\begin{table*}[ht]
\centering
\begin{tabular}{lll}
\toprule
\textbf{Level} & \textbf{Institutional Oversight}  & \textbf{Definition} \\ \midrule
Low & Ad-Hoc Practice &   A mechanism subjectively determined by a developer or user.    \\ 
& Organization Best Practice &  A practice determined by the organization or teams inside of it.    \\  
& Organization Policy  & A policy requiring certain practices to be followed. \\ 
& Industry Standard   & A standard from industry best practice or an official standards body.  \\  
High & Regulation  & The rules and processes necessary to enable a law.  \\  
\midrule
& \textbf{Human Oversight} &  
\\ \midrule
Low & Autonomous AI  & AI output is the final output.     \\ 
& Conditionally Autonomous AI &  AI output could be the final output; otherwise, output is up to the human. \\   
& Human-Approved AI &  AI output must be approved by a human as the final output. \\ 
High & Human-Led with AI-Assistance &  AI output could be used by a human in a larger final output. \\
\bottomrule
\end{tabular}
\caption{Definitions of institutional oversight and human oversight levels from low to high levels of governance strictness.}
\label{table:gov-defs}
\end{table*}

\begin{figure*}[h]
    \centering
\includegraphics[width=0.8\linewidth]{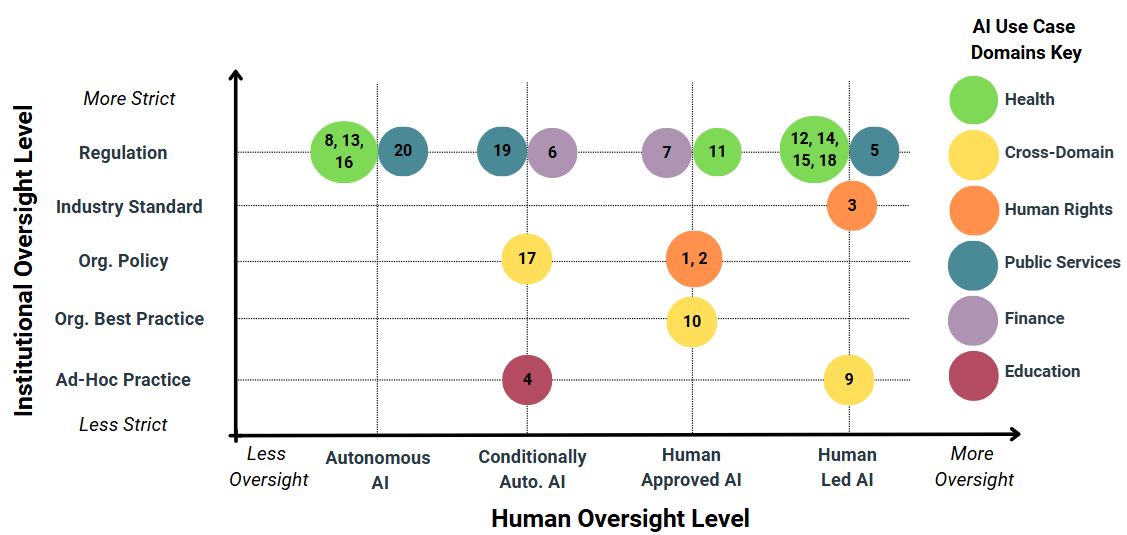}
    \caption{The oversight levels across all of the use cases organized by their domains. The conditionally autonomous AI level is included as “Conditionally Auto. AI” and the human-led with AI-assistance level is written as “Human-Led AI.” We plot the strictest form of institutional oversight for each use case. The numbers match the use case numbers in Appendix B.}
    \label{fig:oversight-by-domain}
\end{figure*}

\subsection{Patterns of AI Governance}
\label{sec:patterns}
From our initial repository of 20 use cases, we expected three main types of AI systems to emerge with varying degrees of human oversight: (1) fully-automated systems where there is no human oversight, (2) human in the loop systems where the human can accept or modify AI outputs, and (3) human in the loop systems where the human can only accept or reject AI outputs, not change them. However, on review of the AI use cases, our analysis resulted in four different levels of governance which we will describe in detail below. We categorize these levels of governance on top of AI systems as types of \textit{human oversight} governance. To determine an AI system's human oversight level, we ask: can a human affect the final output of the system? If so, what actions can they take to determine the final output? 

As we reviewed the mechanisms of governance across use cases, another type of governance emerged; we name this form of governance: institutional oversight. Practitioners shared that institutional oversight can happen internally within AI systems' processes or externally. We define (see Table \ref{table:gov-defs}) and provide examples of the different levels of human oversight and institutional oversight in our initial Fabric repository. We include Figure \ref{fig:oversight-by-domain} which groups each use case by their oversight level and the strictest form of institutional oversight present in the system. For a table of the use cases and their oversight patterns, see Table 2 in Appendix B. 
While the specific governance types and levels we present are not entirely novel, our contribution is in clear labels, consistent structure, and most importantly in documenting the application of governance in deployed AI systems. When referring to the use cases in the following text, we include their number in the repository to find them in Appendix B.

\subsection{Institutional Oversight Levels}

Within the initial Fabric repository, we found multiple governance mechanisms that did not fall under the human oversight levels we uncovered. These institutional oversight mechanisms included a broad range of interventions, such as metric threshold checks, internal policies, industry standards, and regulatory compliance. 
We define the different levels of institutional oversight from less strict governance (e.g., ad-hoc practice) to more strict governance (e.g., regulation) in Table~\ref{table:gov-defs}. We now outline examples of institutional oversight mechanisms from the repository for each governance level. Use cases can include multiple levels of institutional oversight in a single AI system workflow.

\subsubsection{Ad-Hoc Practice}

This level of institutional oversight is the least strict, as individuals or teams decide governance ``off-the-cuff,'' often uninformed by internal guidance or external pressures. Within the repository, two use cases included only ad-hoc practices (the personalized feedback assessor No. 4 and Writing Assistant No. 9) and no stricter levels of institutional oversight were applied. 
\begin{itemize}
\item Personalized Feedback Assessor No. 4: This AI system generates feedback for students based on their online quiz performance. Before sending the feedback to the students, certain safety conditions must be met, including checking if the answer is scientifically correct or if harmful language is used. This safety check is an ad-hoc oversight practice since it is up to the developer to decide what those checks are and what thresholds they should be. If a safety check falls below a certain threshold, the feedback is sent to the student's teacher and the AI system regenerates new feedback. 
\item Writing Assistant No. 9: The organization that provided the personalized writing assistant use case used a third party to procure their AI system. They chose to apply sensitivity filters from the third party to their AI system. As an ad-hoc oversight measure, the organization opted to implement the least restrictive sensitivity filter setting.

\end{itemize}

\subsubsection{Organization Best Practice}
Many practitioners considered their organizational best practices to be an adequate form of institutional oversight. Some of these practices include confidence thresholds set for safety checks (e.g., Insurance Claims Classifier No. 6) and prioritization of certain patients if labeled as high-risk from an AI system (e.g., the CT-Scan Risk Detector No. 14). 
\begin{itemize}
    \item Insurance Claims Classifier No. 6: In the insurance claim workflow, there is a denial confidence threshold check, determined by the insurance organization. This check determines if the case should be considered by a claims analyst or if the AI is allowed to make the final decision.
    \item CT Scan Risk-Detector No. 14: For this example, the AI system flags anomalies of the scan and provides a priority score, after which radiologists see the high-risk scans first. This lets radiologists process next steps for patients who need it most. 
\end{itemize}

\subsubsection{Organization Policy}
Organizations often institute internal policies specific to their sector and domain (e.g., Mental Health Triage Tool No. 13 and OriginTrail Decentralized Knowledge Graph No. 17). If these policies are not followed, then sometimes employees could face repercussions. 
\begin{itemize}
    \item Mental Health Triage Tool No. 13: In the mental health triage tool, we found several examples of governance via internal policy. For a brief insight into the tool, it provides relevant forms for a user to fill out and organizes a preliminary patient report for healthcare providers. The organization policy outlines that a lived-experience digital group must approve a system before it is deployed. A project board, clinical/operational design authority, and technical designer all must sign off before deployment. By including checkpoints, risks are ideally mitigated before deploying systems.
    \item OriginTrail Decentralized Knowledge Graph No. 17: This use case provides knowledge graphs to the public. Users can build their own knowledge graphs. The organization has an open-source policy for their code so that anyone can build knowledge graphs and gain insights from them. Their open-source policy can result in the community flagging safety and risks of their software, supporting governance. 
\end{itemize}

\subsubsection{Industry Standard}
These are standards that could come from industry best practice or a national or international standards body (e.g., International Organization for Standards). Most examples of industry standards came from the healthcare use cases, including medical device standards (e.g., Physio-Risk Reporter No. 8, Mental Health Triage Tool No. 13, and Carer-AI Kit Risk Assessor No. 16). However, other use cases did cover other industry standards; accessibility standards were a priority for a Local Government Chatbot No. 5. 
\begin{itemize}
    \item Local Government Chatbot No. 5: A local government organization built a chatbot for residents to find information about less risky topics on their website. The organization was aware that they needed to make their chatbot accessible for users with visual and other impairments. Given this priority, they followed accessibility standards for their chatbot. 
    \item Mental Health Triage Tool No. 13: This use case provides relevant forms for a user to fill out and organizes a preliminary patient report for healthcare providers. Some of the relevant standards implemented include: safe deployment of healthcare technology with designated safety officers, and a health and equality impact assessment. 
\end{itemize}

\subsubsection{Regulation} 
This level of institutional oversight is the strictest
because regulations are enforceable by law, and failure to comply can lead to legal consequences including sanctions, litigation, or regulatory action

Our use cases captured a broad range of regulations from domains like finance (e.g., Credit Lending Classifier No. 7), healthcare (e.g., Physio-Risk Reporter No. 8), and public services (e.g., Local Government Chatbot No. 5). Almost all of the public sector organization use cases (e.g., Carer-AI Kit Risk Assessor No. 16 and Mental Health Triage Tool No. 13) that process or store sensitive data shared their compliance with data protection regulation as key institutional oversight for their AI systems. We note that how organizations abide by regulation is usually organization dependent.

\begin{itemize}
    \item Credit Lending Classifier No. 6: The AI system in this use case had to comply with data protection regulation since it accesses personal data about credit applicants. Among other practices to comply, this organization had to complete data protection impact assessments.  
    \item Chest X-Ray Abnormality Detector No. 12: This use case involves radiation so must follow radiation regulation. Radiation regulation ensures the safety of patients and those running the scans so is a crucial form of institutional oversight of AI systems.  
\end{itemize}

\subsection{Human Oversight Levels}
We discuss human oversight levels that we observed in the initial Fabric repository. 
Upon conducting our interviews, we noted that our expected ``human in the loop'' AI system types were quite similar and, at times, overlapping. However, we found a key distinction between these two types of AI systems and the human oversight applied---in some instances, the human could update the data given to the AI system to either retrain it or rerun with the same data and system, hoping for different outputs. To account for the nuances in the repository, we redefine the ``human in the loop'' categories into the following human oversight levels: (1) human-approved AI and (2) human-led with AI-assistance. 

We provide specific examples of the human oversight levels observed in Fabric. For each level, we show an example diagram, describe the key abstract features of the oversight level, and discuss at least one of the observed use cases representative of that level.

\subsubsection{Autonomous AI}

\begin{figure}[h]
    \centering
\includegraphics[width=\linewidth]{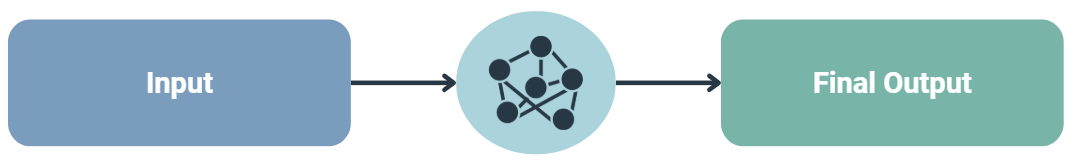}
\caption{The autonomous AI diagram which is the least strict level of human oversight.}

    \label{fig:autonomous-ai}
\end{figure}

Below, we  discuss the autonomous AI governance pattern, see Figure~\ref{fig:autonomous-ai} for a visual.
Features that fall under this human oversight level include:
\begin{itemize}
    \item The AI output is the final output.
    \item No human oversight can change the final output.
\end{itemize}

From the initial Fabric repository, we categorize four use cases as autonomous AI: Physio-Risk Reporter No. 8, Mental Health Triage Tool No. 13, Carer-AI Kit Risk Assessor No. 16, and Image Blurring Tool No. 20. Every autonomous AI use case within Fabric also includes some form of institutional oversight.
\begin{itemize}
    \item  Physio-Risk Reporter No. 8: In this use case, an AI system provides a user (over 65 years old) with a risk score and a report regarding their physio-related status. The user completes prescribed exercises with a physiotherapist (institutional oversight example) while a video is recorded. This video is fed into an AI system, which includes two distinct models before the final output, the risk report, is presented to the user. There is human oversight present for the first AI model but not for the second. A physiotherapist reviews the first models output: a skeletal animation of the user's movements as they complete their exercises. If the skeletal video quality is not adequate, then the physiotherapist can request that the user re-record. We note that there is no governance on the second AI model, the risk report generator. The AI output is the final output aligning this use case with the autonomous AI level. 
\end{itemize}

\begin{figure*}[h]
   \centering
   \includegraphics[width=0.7\linewidth]{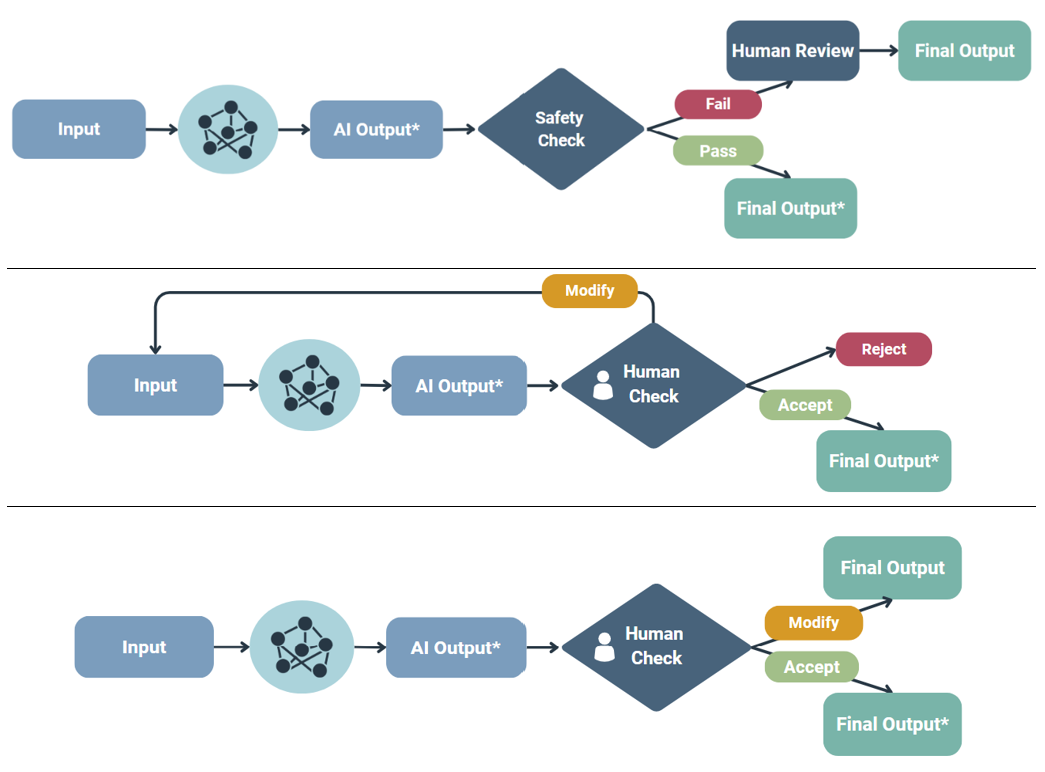}
   \caption{We include three of the human oversight levels in visual form. The top diagram is a conditionally autonomous AI system. The middle diagram is a human-approved AI system. The bottom diagram is a human-led with AI-assistance system. We denote joint human-AI approved final outputs with a *.}
   \label{fig:human-in-ai-loop}
\end{figure*}

\subsubsection{Conditionally Autonomous AI}

For the conditionally autonomous AI oversight pattern, see the top diagram in Figure  \ref{fig:human-in-ai-loop}. The use cases that fall under this human oversight level share the following features:
\begin{itemize}
    \item The AI system can act autonomously. 
    \item If the AI system cannot act autonomously given certain constraints, then the human must make the ultimate decision for the final output. 
\end{itemize}
We observe four use cases in Fabric that are categorized under this oversight level: Personalized Feedback Assessor No. 4, Insurance Claims Classifier No. 6, OriginTrail Decentralized Knowledge Graph No. 17, and Call Center Virtual Assistant No. 19. 
The determining factor for who makes the final decision, the AI or a human, is a safety check which could include a metric or confidence check of the AI output. We note that if the human must participate in the process, we see human oversight like human-approved AI. 

\begin{itemize}
    \item Personalized Feedback Assessor No. 4: The AI system provides students with feedback on their quiz answers. The feedback is generated and given to the student automatically, so the AI system acts autonomously. The AI acts autonomously unless, a teacher has chosen that they want to check the AI feedback before giving it to students. In this instance, we then see a type of human-led with AI-assistance hybrid type of governance, where the teachers accept the feedback or they can edit the AI output before its sent to the student.  
    \item Insurance Claims Classifier No. 6: Unlike the personalized assessment feedback tool where the teacher determines if they want the AI to run autonomously, the insurance claim use case has \textit{two} decision points to determine whether or not the AI system runs autonomously or if human decision-makers need to review the claim. If the claim is simple and the AI system has a high denial confidence (institutional oversight example), then the claim is denied and the AI system has acted autonomously. If the claim is not simple or if the AI system's denial confidence is below a certain threshold, then a claims analyst has to decide if its approved or denied. Whether its an AI or a claims analyst, a customer could appeal the claim as a matter of institutional oversight but this is after the final output.  
\end{itemize}

\subsubsection{Human-Approved AI}
For the human-approved AI governance pattern, see the middle diagram in Figure~\ref{fig:human-in-ai-loop}. 
One of the two human in the loop scenarios must be true for the use case to fall under this human oversight level:
\begin{itemize}
    \item A human can accept the AI output as the final output or reject it (and not have a final output). 
    \item If the human is not satisfied with the AI output, then they can modify the input data to potentially receive a different and better output.
\end{itemize} 
We have five use cases that fall under this level of governance: Metadata Extractor No. 1, PDF Segmentator No. 2, Credit Lending Classifier No. 7, Hyperparameter Optimizer for LLM and RAG Systems No. 10, and Smart Clinical Triage Tool No. 11. 
\begin{itemize}
    \item Metadata Extractor No. 1: The purpose of the AI system is to build a model that extracts user-specified metadata from samples. The user then decides if they are satisfied with the model's predictions and if so, they can use the trained model for their tasks, if not, they can delete the model or update their samples and train a new model. This use case is particularly interesting because it is one of the few uses cases we collected where a trained model is the \textit{final output}. 
    \item Credit Lending Classifier No. 7: The AI system generates a recommendation for credit officers to review before they make the final decision and there are different variables (e.g., loan size) that determine how many credit officers (institutional oversight example) must review the loan application before the final outcome is determined. Ultimately, the credit officers have the final say if they agree or reject the AI system's recommendation. 
\end{itemize}

\subsubsection{Human-Led with AI-Assistance}
For this oversight level see the bottom diagram in Figure \ref{fig:human-in-ai-loop}. In this level, the human always makes the final decision with AI-Assistance. 
The features that fall under this human oversight level include:
\begin{itemize}
    \item The AI system output can be used by the human, if they are satisfied and agree with it.
    \item The final output of the system is more than the AI system output and involves more work by the human. 
\end{itemize}

We have seven use cases that fall under this level of oversight: Family Court Support Chatbot No. 3, Local Government Chatbot No. 5, Writing Assistant No. 9, Chest X-ray Abnormality Detector No. 12, CT Scan Risk Detector No. 14, Consultation AI Note-Taker No. 15, and Automated Imaging Protocol Selector No. 18. 

At this level of human oversight, the human can decide to keep working with the support of the AI (e.g., Family Court Support Chatbot No. 3) or the human can decide to use the AI output for their final decision (e.g., Chest X-Ray Abnormality Detector No. 12, and CT-Scan Risk Detector No. 14). 

\begin{itemize}
    \item Family Court Support Chatbot No. 3:  The human can keep chatting with the chatbot tool until they either have their answer or decide to take their query to a human for support. This use case has loops in the functionality so the human can keep requesting things of the AI system.
    \item CT Scan Risk Detector No. 14: The radiologist who views the initial scan and receives the secondary capture from the AI system makes uses of their own knowledge and the AI  output if they agree with it in their final report; hence the AI assistance is there if deemed useful by the human. 
\end{itemize}

\section{Discussion}
\label{sec:disc}

We outline the main takeaways from the curation and analysis of the initial Fabric repository. We also discuss the value and limitations of our particular methodology. 

\subsection{Patterns of Oversight}
The levels of human oversight and institutional oversight discussed above, together, make a ``Pattern of Governance'' and more specifically, a ``Pattern of Oversight.'' 
Through our study, we saw these ``Patterns of Oversight'' which are the decisions of the two levels of oversight paired together. We argue that these patterns should be included in model cards and documentation for transparency purposes. Currently, much of the thinking behind governance happens post-development of the AI system but pre-deployment. We encourage the community to discuss these patterns pre-development and to revisit them continuously throughout the AI life cycle. In Appendix C, we include a questionnaire to support individuals or teams to check what level of human oversight their AI use case falls under; in addition, we provide some questions to support discussions as to what institutional oversight might be necessary given different risks.

We outline common patterns of oversight that we see in our repository below. We note that many of our use cases are considered high-risk in line with the EU AI Act's framework; high-risk systems are those used within contexts where failure can significantly harm health, safety, or fundamental rights \cite{euai2024}. It is therefore unsurprising that many of them followed the strictest level of institutional oversight. 
\begin{itemize}
    \item The most common human oversight pattern is ``human-led with AI assistance,'' so a decision-maker can modify or accept an AI's output as a final output. The most frequent level of institutional oversight was regulation for these use cases, which was also the most common institutional oversight across all use cases.
    \item All AI use cases that fall into the autonomous AI oversight level abide by the strictest form of institutional oversight, regulation.
    \item All conditionally autonomous AI use cases, except for the more high-risk financial use case, follow the least strict level of institutional oversight: ``ad-hoc practice.'' 
\end{itemize}

\subsection{Institutional Oversight Considerations}

The majority of the practitioners that we interviewed for our repository could not point to best practice AI governance guidelines. This lack of clarity on institutional oversight could reflect either an absence of sufficient governance, or a design flaw in the governance process that prevents actors within the organization from being aware. In both cases, the use of AI is not effectively guided or constrained. However, practitioners could point to regulation or domain-specific industry standards that they followed. The majority of use cases included at least some form of regulatory compliance in their AI systems. This was especially clear for government use cases. In less risky domains, ad-hoc practices were common while stricter levels of governance were often absent, suggesting that practitioners may have struggled to integrate emerging AI governance and safety measures alongside industry requirements.

We found that most of the institutional oversight was decided in-house to comply with regulations, industry standards, organizational policies, or general ``best practices.'' Some practitioners acknowledged that while they did not have specific AI governance in place when the AI systems were built, they are working on or have governance frameworks in place internally now. This was the case for the Local Government Chatbot No. 5 use case. This finding highlights how the practitioners across industries (at least in our recruited pool) are aware of the risks posed by using AI systems in their processes and are prioritizing best practices or policy frameworks internally for current and future AI use once the technical envelopment of their use cases are production/user ready.

Many public and private sector use cases leveraged AI systems that were not developed internally but were procured from third-parties. Those deploying third-party systems had less control of system innards. The choice to use third-party AI systems stems from the appeal of accessibility and ease of use. Perhaps more concerning, there is a lack of visibility across organizations that may make attempts to debug, escalate, and correct particular failures challenging. 

Some interviewed practitioners expressed the need for more knowledge sharing of what AI governance can look like \textit{across} industries and sectors. They were curious to see how others had implemented AI systems and included forms of governance. This interest from the practitioners themselves points to the fact that the Fabric repository resonates and will provide valuable examples of governance in action.

In practice, institutional oversight still varies widely. There are several reasons for this, including:
\begin{itemize}
    \item The locus of control (primarily in either the Chief Tech Officer or Chief Risk Officer) is not standardized, often resulting in misaligned governance.
    \item International, national, and local legislation, which can differ significantly, is evolving (or in some cases is completely absent), which creates a lack of clarity and a hesitance to create heavy-weight processes that might not respond to regulatory shifts.
    \item There has yet to be a body of generally accepted practices that has been developed around AI, as the field is in its infancy, creating even more uncertainty in the space.
\end{itemize}
Fabric may provide support to facilitate alignment, clarity, and, in future iterations, establish best practices.

\subsection{Human Oversight Considerations} 
As noted before, all four of the autonomous AI use cases follow the strictest level of institutional oversight (see Figure 2 above, and Table 2 in Appendix B). This is notable because these systems are subject to greater levels of regulatory requirements, suggesting they may be higher risk, yet they have the least amount of human oversight. For the conditionally autonomous AI systems (when the output is not fully autonomous), we see the human-approved AI pattern appear in all use cases. The organizations we survey recognize that there can be a combination of human oversight patterns within a system process. 
Most of the use cases in Fabric fall under the human-led AI assistance human oversight category. This category includes the highest level of human oversight with an AI system still ``in the loop.'' In this instance, the human can decide if the AI output should contribute to the final output, which requires human labor beyond the potential AI contribution. Although this potentially requires more labor from the human involved, such a governance structure can allow for a strong safeguard before integrating, if at all, the AI output into a decision~\cite{steyvers2022bayesian,collins2024modulating}. 
We highlight that human oversight levels give rise to a set of controls that must be followed; the deciding factor of what level is necessary often depends upon the risk rating of the use case.

We note that the human could be over- or under-reliant on the AI system and result in negative outcomes. Specifically, the concept of automation bias, when users are prone to accept AI-generated outputs without adequate critical evaluation, and the harms are well-documented \cite{skitka1999does, gosline2024nudge}. The consequences of over-reliance on AI can also negatively impact the governance of such systems ~\cite{sauer2012comparison}.
For instance, if the human over-relies on the system and does not adequately check the output for errors, they could include a mistake in the final output, like leading to a misdiagnosis of a patient~\cite{logg19people}. On the other hand, the AI system could provide a useful output like a critical anomaly in a chest X-ray that the doctor had missed; the doctor could completely overlook the AI's output because they under-rely on the system, resulting in a less robust final report being produced~\cite{bonaccio_dalal}. Fabric use cases that fall under the human-led AI assistance and human-approved AI levels are common in the literature when studying what sorts of human oversight or technical guardrails could be useful to apply for less bias or uncertain outcomes~\cite{hou2021expert,lai2021towards}.

\subsection{Values and Limitations of Our Study}

While we cannot conclude anything about the effectiveness of human oversight and the downstream impacts from an analysis of Fabric alone, we argue that focusing on oversight is crucial to understand how governance looks like in practice. Future research must study the effectiveness of each mechanism in our repository. We hope to then link these results in our project website\footnote{ \url{https://fabric-repository.github.io}} later to provide more insights for practitioners. 

An inherent value in our approach is that we used qualitative methods for data collection, allowing for more unique insights to emerge. For instance, some practitioners did not initially consider how their AI system included multiple ML models until midway into the interview (e.g., Physio-Risk Reporter No. 8 and PDF Segmentator No. 2). 
Some practitioners seemed to hold an abstract view of AI in organizational processes and did not consider the fine-grained specification of the AI system as a whole. This points to differences in the practitioner perception of AI versus its actual role within the workflows, and the value of open-ended interviewing and diagram development for the discovery and nuance of oversight.  

We also highlight that co-designing use case diagrams in the interviews was an effective practice to open up the discussion about the overall system. We found that the diagram discussions led to the best understanding of the human oversight and institutional oversight patterns. The more interviews we conducted, the more attuned we were to know what kind of questions answered and data we needed to build the repository and analyze the oversight patterns. 

Since we used our personal and professional networks to recruit practitioners as participants, we encounter selection bias in our recruitment; although, we attempted to mitigate some of this concern with snowball sampling. In future iterations of Fabric, we hope to recruit a more representative sample. We see the double-edged sword of our semi-structured interviews; while such personal conversations are highly information-rich, they are difficult to scale. 

In next iterations of Fabric, we plan to update our interview questions to include more specific questions about third-party AI systems and different aspects of institutional oversight. The earlier interviews we ran were less focused on institutional oversight, as this type of governance became more apparent later on in our interviewing. To mitigate this discrepancy, we followed up with practitioners and received varying levels of detail. We also note that the oversight mechanisms are often domain-specific; as such, domain-specific findings may become clearer in future iterations of Fabric. As AI systems themselves adapt, we anticipate continually revisiting and adapting Fabric, particularly as new levels or patterns of oversight emerge. 
Fabric could also be married with other approaches for tracking changes in AI systems like FeedbackLogs~\citep{barker2023feedbacklogs}.

\section{Conclusion}
\label{sec:conc}
In this paper, we release Fabric, a public repository of AI use cases and corresponding governance. We collect examples of deployed AI systems from practitioners through semi-structured interviews. In these interviews, we co-design diagrams of a practitioner's AI system and corresponding governance. Through these interviews, we aimed to understand oversight is implemented in practice and where these mechanisms lie within AI workflows. This knowledge fills a crucial gap in the community's knowledge base. We discuss two types of oversight that emerge from the use cases that we collate as part ofy Fabric: human oversight and institutional oversight. Human oversight relates to how much oversight of the AI system occurs immediately before a final output is determined. Institutional oversight is linked to interventions that occur along the system workflow that add layers of safety to ensure the reliability and trustworthiness of the system. 

From these interviews, we curate and release Fabric, a deployed AI use case repository, that leverages easy-to-interpret diagrams to characterize how AI systems are governed by different levels of oversight. We release 20 initial use cases in the repository. 
In the next iteration of Fabric, we will add functionality to the repository for other practitioners to add their own use cases. This will allow Fabric to grow. As we collect more use cases, we garner more insights about AI use at scale in the real-world. We hope to see other patterns of oversight arise in future iterations of Fabric. Crucially, we will begin assessing the efficacy of different governance mechanisms on target outcomes. 
By growing our repository, we can foster broader societal and academic awareness of the ever-evolving use of AI in the \textit{fabric} of our everyday life.

\section*{Acknowledgments}
The work of MJ, KB, UB, and AW is supported in part by ELSA - European Lighthouse on Secure and Safe AI funded by the European Union under grant agreement No. 101070617. Views and opinions expressed are however those of the author(s) only and do not necessarily reflect those of the European Union or European Commission. Neither the European Union nor the European Commission can be held responsible for them. AW also acknowledges support from a Turing AI Fellowship under EPSRC grant EP/V025279/1, The Alan Turing Institute, and the Leverhulme Trust via CFI. KMC acknowledges support from the Cambridge Trust and King's College Cambridge. We would like to thank Diletta Huyskes for her support in the early stages of the project and Sophia Worth for her feedback on the paper. We also greatly appreciate the contributions that Michael Convey, Runcong Zhao, Jurij Skornik, Michelle James, Sameer Gangoli, Koen Cobbaert, Jenny Partridge, Enrico Ferraris, and Dr. Elango Vijaykumar (Modality Partnership) made for Fabric.  

\section*{Ethical Considerations Statement}
While we foresee ample positive benefits of Fabric based on the practitioners we interviewed, we understand that there are privacy risks, some of which we aimed to mitigate by keeping the organization and practitioner name anonymous (unless the practitioner specifically gave their consent and requested for their organization to be de-anonymized). We did make the risk clear to the practitioners that they could be re-identified from the repository, even though we took necessary precautions. Though we do not suspect this, practitioners could have mislead us about the governance that they have in place; to some extent, regulatory compliance is key, so most organizations want to project strength and purport adherence. That said, we have no way of verifying or auditing the correctness of a practitioner's claims, but do have reviewed sign-offs for each of the use cases.

Although not a focus of our study, people affected by the AI systems we analyze will benefit from gaining an increased awareness of what guardrails and oversight are included in AI systems. There is no direct risk from our research for the people affected by AI systems in our repository; we do not include any specific instances of them or have access to any of their data. 

In addition, AI practitioners (those we have not met with), by reviewing our repository, can better understand how governance can look like in practice and implement it accordingly for their own AI use cases. However, there is a risk that they over-rely on the repository instead of critically thinking about what governance mechanisms make the most sense or will be most effective for their use case.

\clearpage

\appendix

\renewcommand\thesection{\Alph{section}}
\graphicspath{{Fig-Use-Cases/}{Other-Figs/}}

\input{appendix}

\end{document}

%% file: appendix.tex
\onecolumn

\section{Interview Questions}

We ask the following questions in our semi-structured interviews to collect the initial Fabric repository. Practitioners often answered multiple questions in their responses. The conversation remained open ended to allow for natural elaboration and feedback. 

\noindent\begin{enumerate}[itemsep=0.25em, topsep=0.25em]
    \item System-Level Questions
    \begin{enumerate}[itemsep=0.25em, topsep=0.25em]
        \item Do you have a specific AI use case in mind for our conversation?
        \item What task does the AI system complete or help complete? How do they support in this task?
        \item How would you categorize the AI system (e.g., data driven, including subsymbolic AI or machine learning, or rules/logic-driven, including symbolic AI and)?
        \item Does the AI system operate autonomously or collaboratively?
        \item How do users interact with the system? Can they edit, modify, or override outputs?
        \item What is the AI system's output?
    \end{enumerate}
    \item Governance Questions
    \begin{enumerate}[itemsep=0.25em, topsep=0.25em]
        \item Who makes the final decision, the AI system or the user?
        \item What external regulations or internal policies govern the AI system's use?
        \item How are risks and compliance assessed?
        \item In deciding what mechanisms of governance to include in your system, did you consider any best practice guidance? If so, what frameworks or guidance did you take inspiration from?   
    \end{enumerate}
    \item Impact Questions
    \begin{enumerate}[itemsep=0.25em, topsep=0.25em]
        \item What were the intended impacts of the system? 
        \item What impacts has the AI system had?
        \item How are the impacts measured?
    \end{enumerate}
\end{enumerate}

\newpage

\section{Fabric Repository: Use Cases}
In this appendix, we (1) include a summary statistics section of all of the use cases, and (2) include the full repository. 
\label{app:diagrams_all}
\begin{table}[H]
\centering
\resizebox{\linewidth}{!}{%
\begin{tabular}{l l c c c c c}
\toprule
\textbf{Use Case Name \& Number} & \textbf{Human Oversight} & 
\textbf{Ad-Hoc} & \textbf{Org. Best} & 
\textbf{Org. Policy} & \textbf{Industry Std.} & \textbf{Regulation} \\
\midrule
Physio-Risk Reporter No. 8 & Autonomous AI &  & \checkmark & \checkmark &  & \checkmark \\
Mental Health Triage Tool No. 13 & Autonomous AI &  & & \checkmark & \checkmark & \checkmark \\
Carer-AI Kit Risk Assessor No. 16 & Autonomous AI &  &  &  & \checkmark & \checkmark \\
Image Blurring Tool No. 20 & Autonomous AI & & & \checkmark & & \checkmark \\

Personalized Feedback Assessor No. 4  & Conditionally Autonomous AI &  \checkmark &  &  &  &  \\
Insurance Claims Classifier No. 6 & Conditionally Autonomous AI &  & \checkmark & \checkmark &  & \checkmark \\
OriginTrail Decentralized Knowledge Graph No. 17 & Conditionally Autonomous AI & \checkmark &  & \checkmark &  &  \\
Call Center Virtual Assistant No. 19 & Conditionally Autonomous AI & & \checkmark & \checkmark & \checkmark & \checkmark \\

Metadata Extractor No. 1 & Human-Approved AI &  \checkmark & \checkmark & \checkmark &  &  \\
PDF Segmentator No. 2 & Human-Approved AI &  \checkmark & \checkmark & \checkmark &  &  \\
Credit Lending Classifier No. 7 & Human-Approved AI &  &  & \checkmark &  & \checkmark \\
Hyperparameter Optimizer for LLM \& RAG Systems No. 10  & Human-Approved AI &  &  \checkmark &  &  & \\
Smart Clinical Triage Tool No. 11 & Human-Approved AI & \checkmark & \checkmark &  \checkmark &  \checkmark &  \checkmark \\

Family Court Support Chatbot No. 3  & Human-Led w/ AI Assist. &  &  &  & \checkmark &  \\
Local Government Chatbot No. 5 & Human-Led w/ AI Assist. &  &  & \checkmark &  & \checkmark \\
Writing Assistant No. 9 & Human-Led w/ AI Assist. &  \checkmark &  &  &  &  \\
Chest X-Ray Abnormality Detector No. 12 & Human-Led w/ AI Assist. &  & \checkmark &  & & \checkmark \\
CT Scan Risk Detector No. 14 & Human-Led w/ AI Assist. &  & \checkmark & \checkmark &  & \checkmark\\
Consultation AI Note-Taker No. 15 & Human-Led w/ AI-Assist. &  &  &  & \checkmark & \checkmark \\
Automated Imaging Protocol Selector No. 18 & Human-Led w/ AI-Assist. & & \checkmark & & \checkmark & \checkmark \\
\bottomrule
\end{tabular}%
}
\caption{We provide the levels of oversight across the repository, with the human oversight level and institutional oversight levels: ad-hoc practice, organization best practice, organization policy, industry standard, and regulation. The use case numbers align with their ordering in the repository.}
\label{tab:inst-governance-fit}
\end{table}

\subsection{Summary Statistics}
\begin{table}[hb!]
\centering
\scriptsize
\begin{tabular}{@{}ll@{}}
\toprule
\textbf{Category} & \textbf{Count} \\ \midrule
\multicolumn{2}{@{}l}{\textbf{Organizational Sector}} \\
Public& 12\\
Private& 6\\
Civil Society& 2\\ \midrule

\multicolumn{2}{@{}l}{\textbf{System Type}} \\
Language Model & 11\\
Non-Language Model & 9\\ \midrule

\multicolumn{2}{@{}l}{\textbf{Domain}} \\
Health& 8\\
Cross-Domain& 3\\
Finance& 2\\
 Human Rights&3\\
 Education&1\\
 Public Services&3\\ \midrule

\multicolumn{2}{@{}l}{\textbf{Repeated Queries of the System}} \\
Multiple queries/ongoing use & 9\\ 
Single query & 11\\ \midrule

\multicolumn{2}{@{}l}{\textbf{Architecture Features}} \\
Uses multiple models& 2\\
Non-static (retrainable) & 1\\ 
Other& 17\\ \midrule

\multicolumn{2}{@{}l}{\textbf{Types of Data}} \\
Personal and/or sensitive data & 14\\ 
Non-personal nor sensitive data & 6\\ \midrule

\multicolumn{2}{@{}l}{\textbf{Human Oversight}} \\
Autonomous AI & 4 \\
Conditionally autonomous AI & 4\\
Human-approved AI & 5\\
Human-led with AI assistance & 7\\
\bottomrule
\end{tabular}
\caption{We show summary statistics from the initial Fabric repository of 20 AI use cases.}
\label{tab:fabric-summary}
\end{table}

Table~\ref{tab:fabric-summary} summarizes the distribution of use cases across key attributes. We observed a variety of organizational sectors, domains, technical architectures, and human oversight levels. Given the rise of large language models (LLMs) in the AI industry, we distinguish our AI systems collected by language models (e.g., LLMs and traditional natural language processing models) versus non-language models (e.g., computer vision models and classifiers).   
We also categorized each AI system based on how often systems are called within a single process workflow (e.g., chatbots can be queried multiple times but classification systems are typically queried only once) to potentially support ongoing interactions. The majority of use cases involved a single AI system query. Some use cases rely on multiple AI models working in sequence or parallel within a workflow. 
We categorize each use case according to the degree of human oversight at the point where the AI system's output could be used in a final output of the process. We also include Table \ref{tab:inst-governance-fit} of the AI use cases to understand the different levels of oversight that the use cases follow.

\subsection{All Use Cases}
\begin{figure*}[h!]
    \centering
\includegraphics[width=0.9\linewidth]{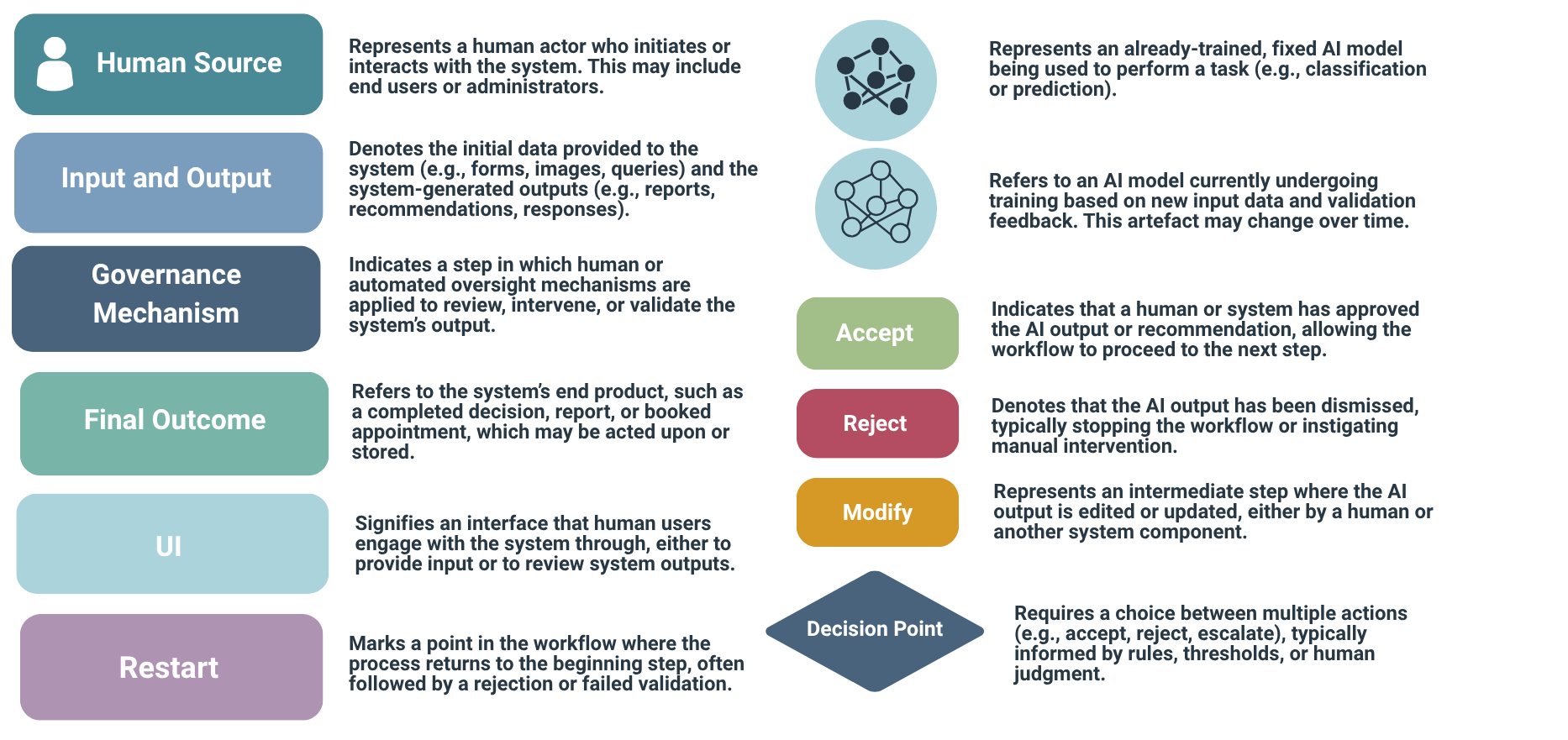}
    \caption{We outline the legend for the AI workflow diagrams.}
    \label{fig:diagram-key}
\end{figure*}

We include details about each of the use cases in the initial Fabric repository.
For a legend of the AI workflow diagrams, see Figure~\ref{fig:diagram-key}. \\
\clearpage
\newpage

\noindent\begin{usecase}{1}{Metadata Extractor}{Civil Society}{Human Rights}{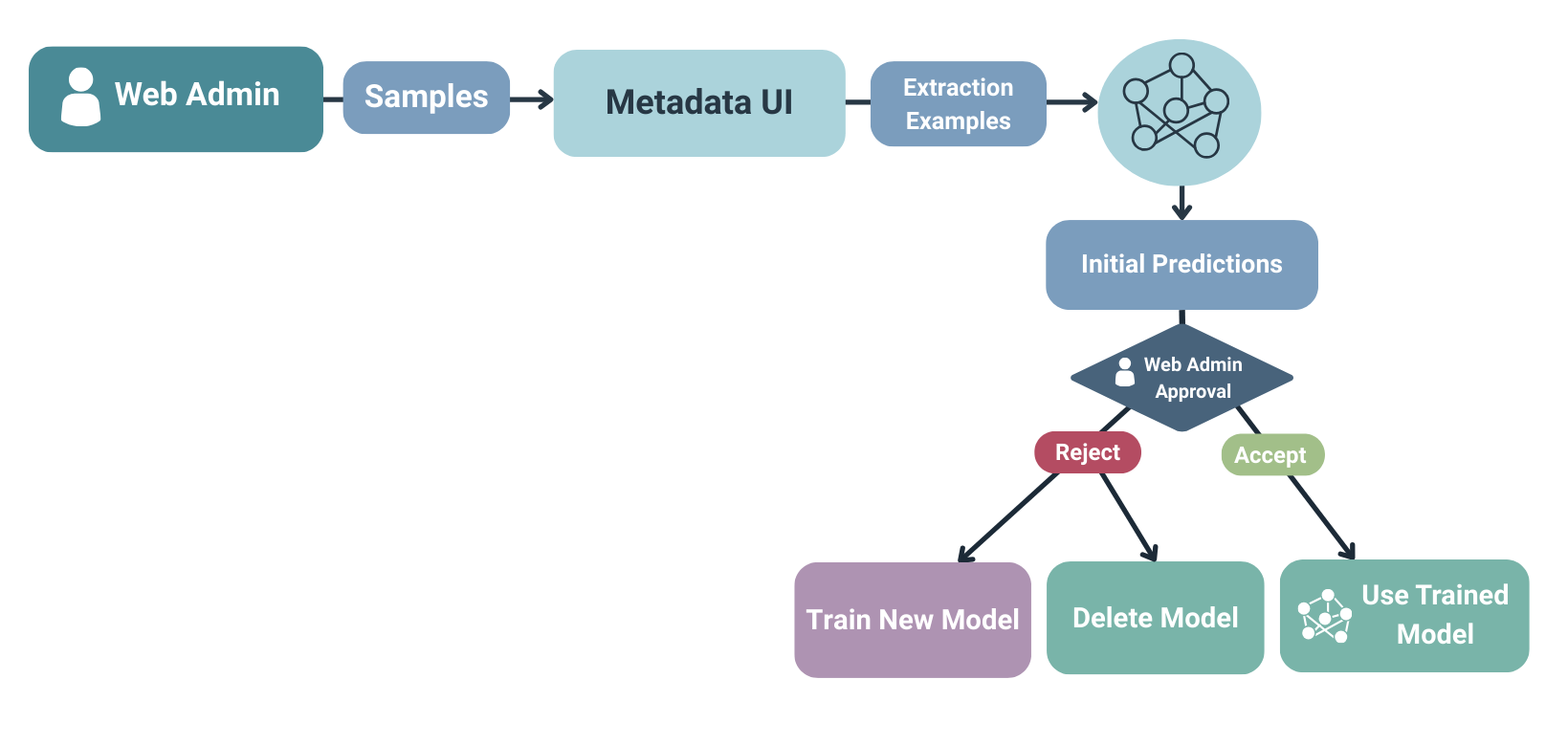}{fig:metadata}
\textbf{Task:} A user can train an AI model to extract metadata from text or documents. If the user is satisfied with the performance of the outputted model, then they can use it.\\
\textbf{Intent:} The main impact of this AI system is to save time for users who cannot afford to scan through hundreds of PDFs or paragraphs. The AI supports efficiency.\\
\textbf{Risks:} The trained AI model could make mistakes when the user uses it.\\
\textbf{Human Oversight Level:} Human-Approved AI\\
\textbf{Institutional Oversight Examples:} Models should be pulled from a specific commit number (ad-hoc practice), services that use the AI should have a release version (ad-hoc practice), all benchmarks should be saved in a public repository (organization best practice), test sets to assess that performance is maintained (organization best practice), code implemented is open-sourced (organization policy), services should be covered with unit tests, integration tests, and end-to-end tests (organization policy)\\
\textbf{Explanation of AI Workflow:}\\
\textit{Input}: The user submits samples to the metadata UI, which then extracts the necessary information to train an AI model.\\
\textit{Process}: The AI model generates its initial predictions based on the metadata structure given to the system.\\
\textit{Output}: The user reviews the newly trained AI model's predictions and either accepts (and finalizes the model) or rejects the model (and deletes or trains again with updated data).\\
\end{usecase} \\
\newpage

\noindent\begin{usecase}{2}{PDF Segmentator}{Civil Society}{Human Rights}{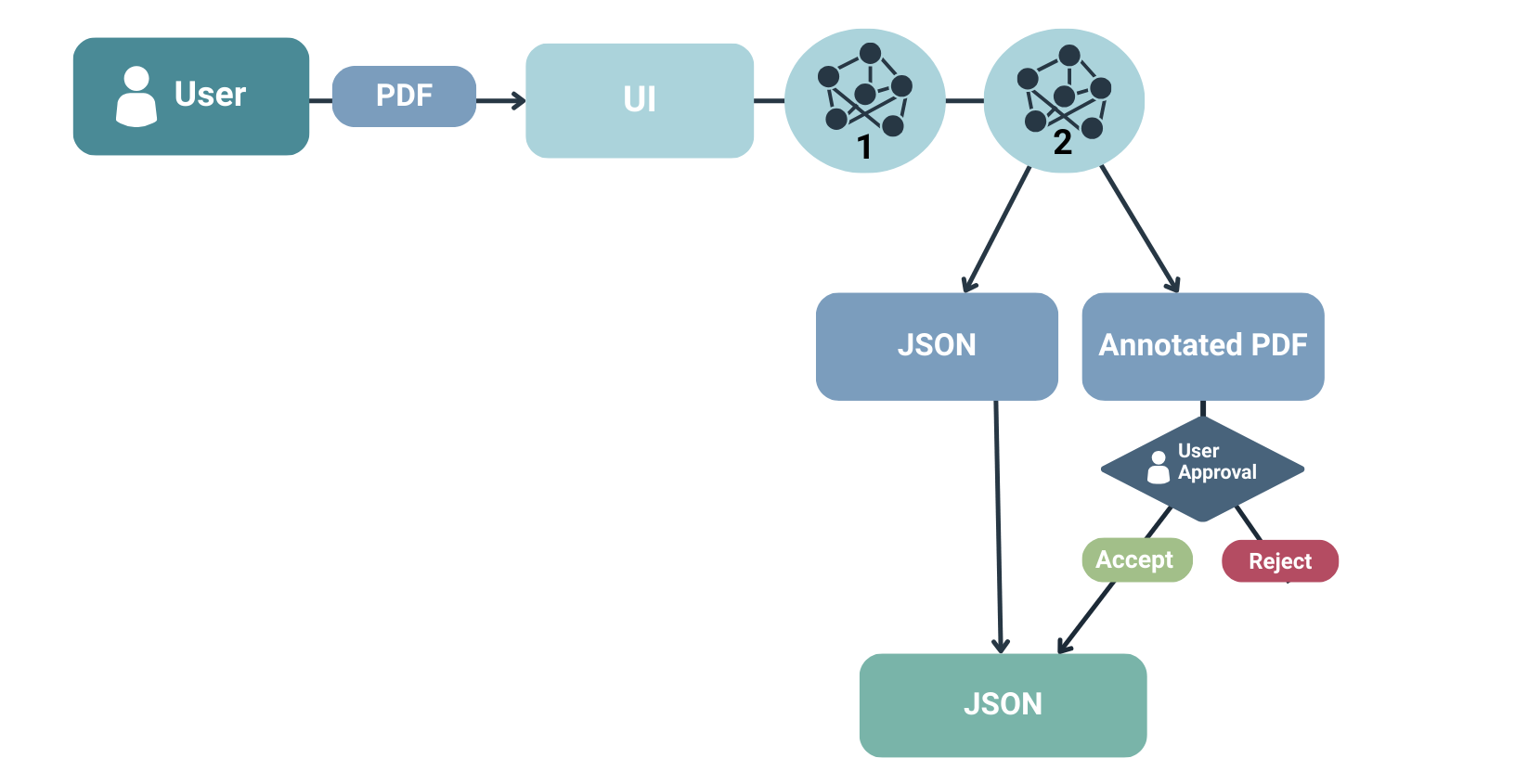}{fig:segmentator}
\textbf{Task:} A user requires their PDF to be segmented, so an AI system segments it into an annotated PDF and JSON.\\
\textbf{Intent:} The AI system supports users in collecting information they need for different tasks.\\
\textbf{Risks:} The AI system could make mistakes when annotating and segmenting the PDF.\\
\textbf{Human Oversight Level:} Human-Approved AI\\
\textbf{Institutional Oversight Examples:} Models should be pulled from a specific commit number (ad-hoc practice), services that use the AI should have a release version (ad-hoc practice), all benchmarks should be saved in a public repository (organization best practice), test sets to assess that performance is maintained (best practice), code implemented is open-sourced (organization policy), services should be covered with unit tests, integration tests, and end-to-end tests (organization policy)\\
\textbf{Explanation of AI Workflow:}\\
\textit{Input}: A user submits a single PDF to a user-interface (UI).\\
\textit{Process}: Model 1 segments the PDF; its output feeds Model 2 which classifies the segmentations. Outputs are a JSON and an annotated PDF.\\
\textit{Output}: The user checks the output. If satisfied with the annotated PDF, they accept it and use the JSON; otherwise, they reject and do not use the JSON.\\
\end{usecase} \\
\newpage

\noindent\begin{usecase}{3}{Family Court Support Chatbot}{Public}{Human Rights}{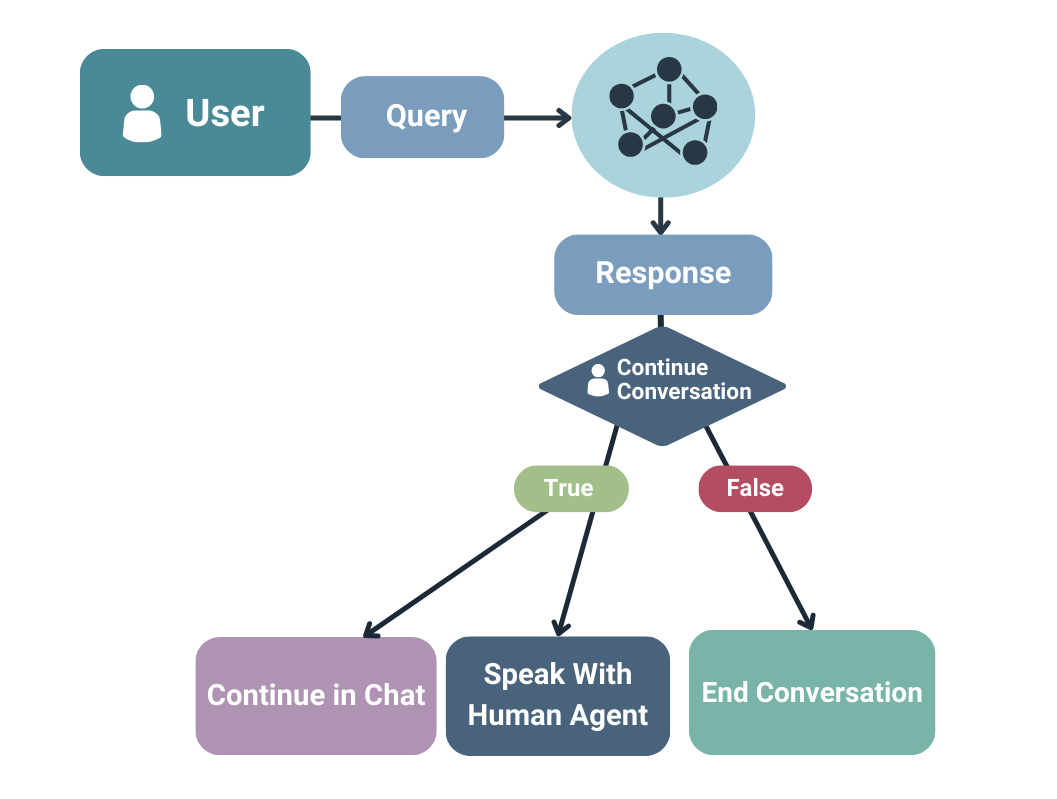}{fig:court}
\textbf{Task:} This AI system generates responses to questions that users ask in a chat format. The AI pulls from information from the institution's website and can route users to speak with a human representative if required.\\
\textbf{Intent:} The AI system provides immediate, around-the-clock responses and reduces the workload of employees.\\
\textbf{Risks:} The AI system could make incorrect or poor-quality responses, which in turn could misinform users. Sensitive information shared in the chat may not be completely secure in extreme instances.\\
\textbf{Human Oversight Level:} Human-Led with AI-Assistance\\
\textbf{Institutional Oversight Examples:} AI cyber-security standard (industry standard), algorithmic transparency standard (industry standard)\\
\textbf{Explanation of AI Workflow:}\\
\textit{Input}: A user submits a query to the AI system.\\
\textit{Process}: The AI system generates a response to the user in the chat. The user views the response and decides if they want to continue the chat.\\
\textit{Output}: If the user wants to continue the chat but speak to a human agent, they can be routed to an employee who can view their chat history and support them. If the user does not want to continue the chat, they can exit and end the chat.\\
\end{usecase}\\
\newpage

\noindent\begin{usecase}{4}{Personalized Feedback Assessor}{Public}{Education}{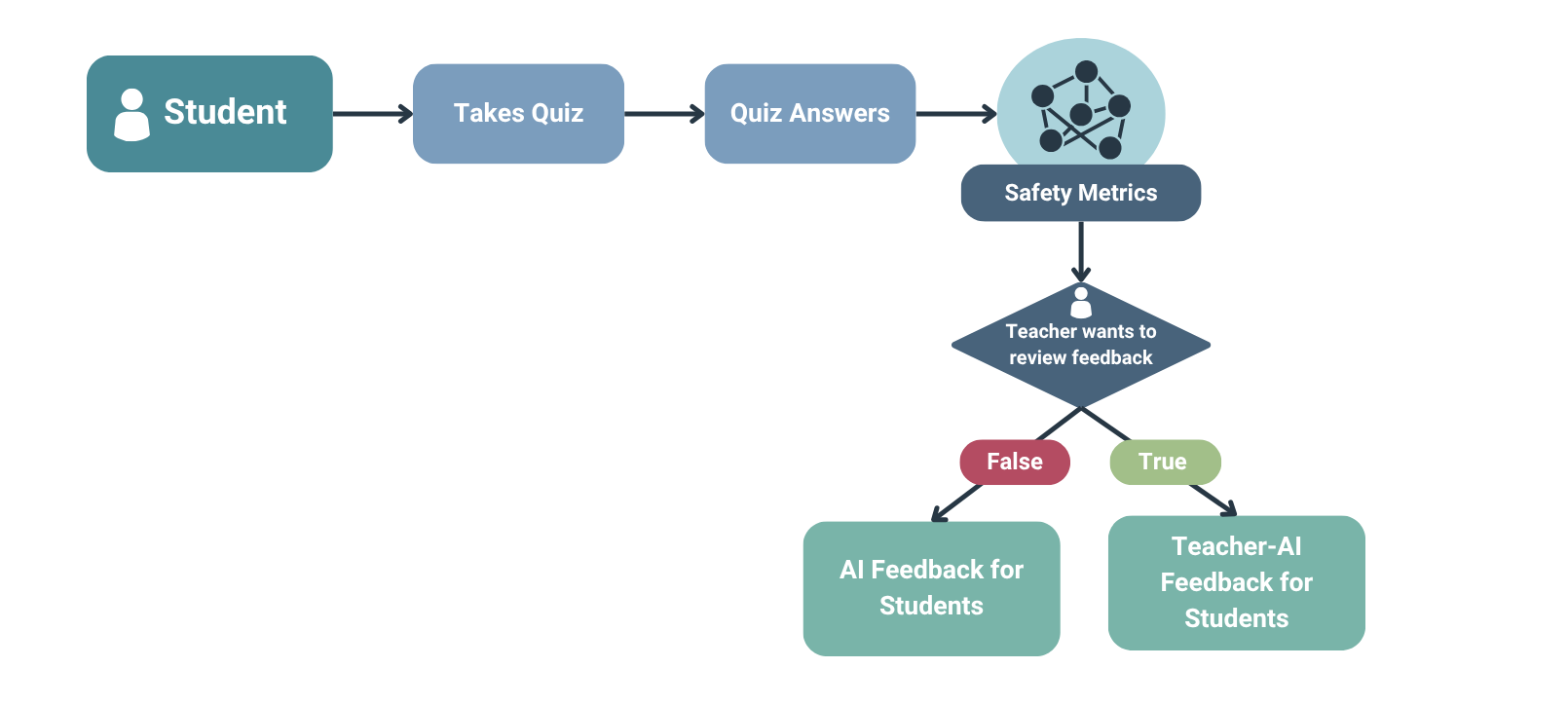}{fig:assessment-tool}
\textbf{Task:} Students take practice quizzes to prepare for their exams and this AI system supports them by providing personalized feedback on their answers. \\ 
\textbf{Intent:} The AI system gives students personal feedback to support and speed up their learning process; it can also save their teacher's time on marking.\\
\textbf{Risks:} The AI system can make mistakes which could confuse students.\\
\textbf{Human Oversight Level:} Conditionally Autonomous AI\\
\textbf{Institutional Oversight Examples:} Safety metrics and chosen safety thresholds (ad-hoc practice)\\
\textbf{Explanation of AI Workflow:}\\
\textit{Input}: A student takes a quiz and the quiz answers are passed to the AI system.\\
\textit{Process}: The AI system reviews answers and generates feedback. The feedback is checked with safety metrics (e.g., scientific correctness, harmful language, alignment to student knowledge). If the score is below a threshold, feedback is sent to the teacher and regenerated.\\
\textit{Output}: Once feedback passes safety metrics and/or teacher approval, it is sent to the student.\\
\end{usecase}
\newpage

\noindent\begin{usecase}{5}{Local Government Chatbot}{Public}{Public Services}{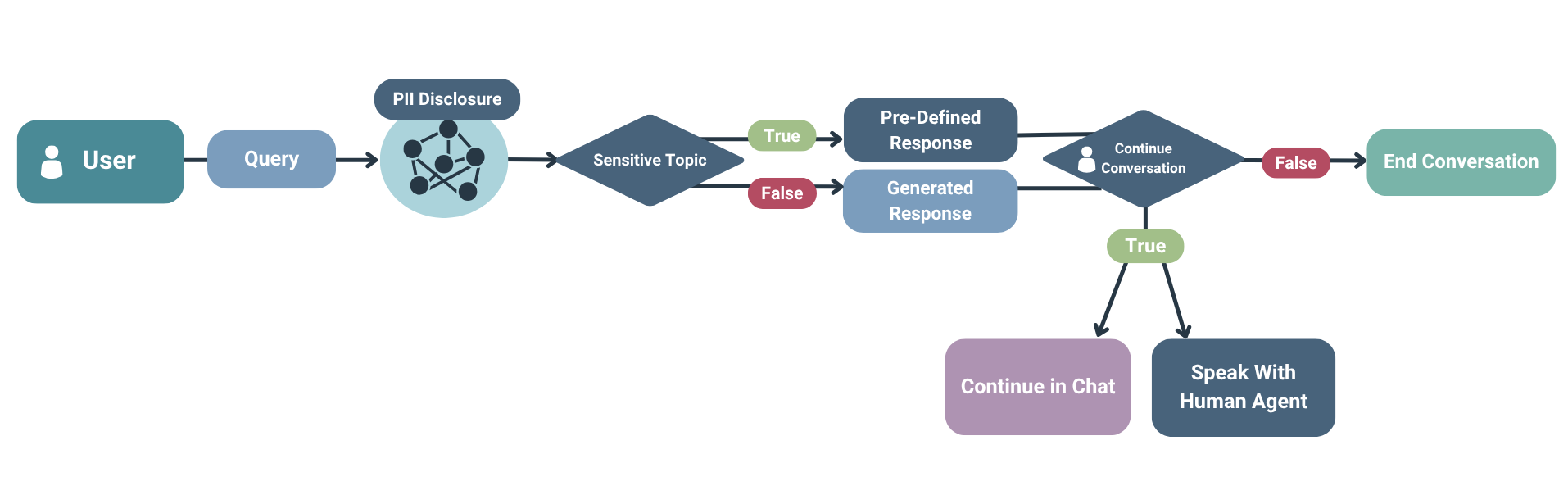}{fig:localgov}
\textbf{Task:} A user queries a chatbot and asks questions about information on a website.\\
\textbf{Intent:} The AI system provides efficient 24/7 answers and reduces administrative burden from simple, non-personal questions.\\
\textbf{Risks:} Despite warnings, users might submit personally identifying information (PII); incorrect or misinterpreted responses could misinform users.\\
\textbf{Human Oversight Level:} Human-Led with AI-Assistance\\
\textbf{Institutional Oversight Examples:} Internal digital board (organization policy), data privacy (organization policy), data protection (regulation), accessibility requirements (regulation)\\
\textbf{Explanation of AI Workflow:}\\
\textit{Input}: A user (instructed not to provide PII) submits a query.\\
\textit{Process}: The AI system decides if the topic is sensitive. If sensitive, it outputs a pre-defined response; if not, it generates a response from the website content knowledge base.\\
\textit{Output}: The user can continue asking questions, end the conversation, or be directed to a human agent.\\
\end{usecase}
\newpage

\noindent\begin{usecase}{6}{Insurance Claims Classifier}{Private}{Finance}{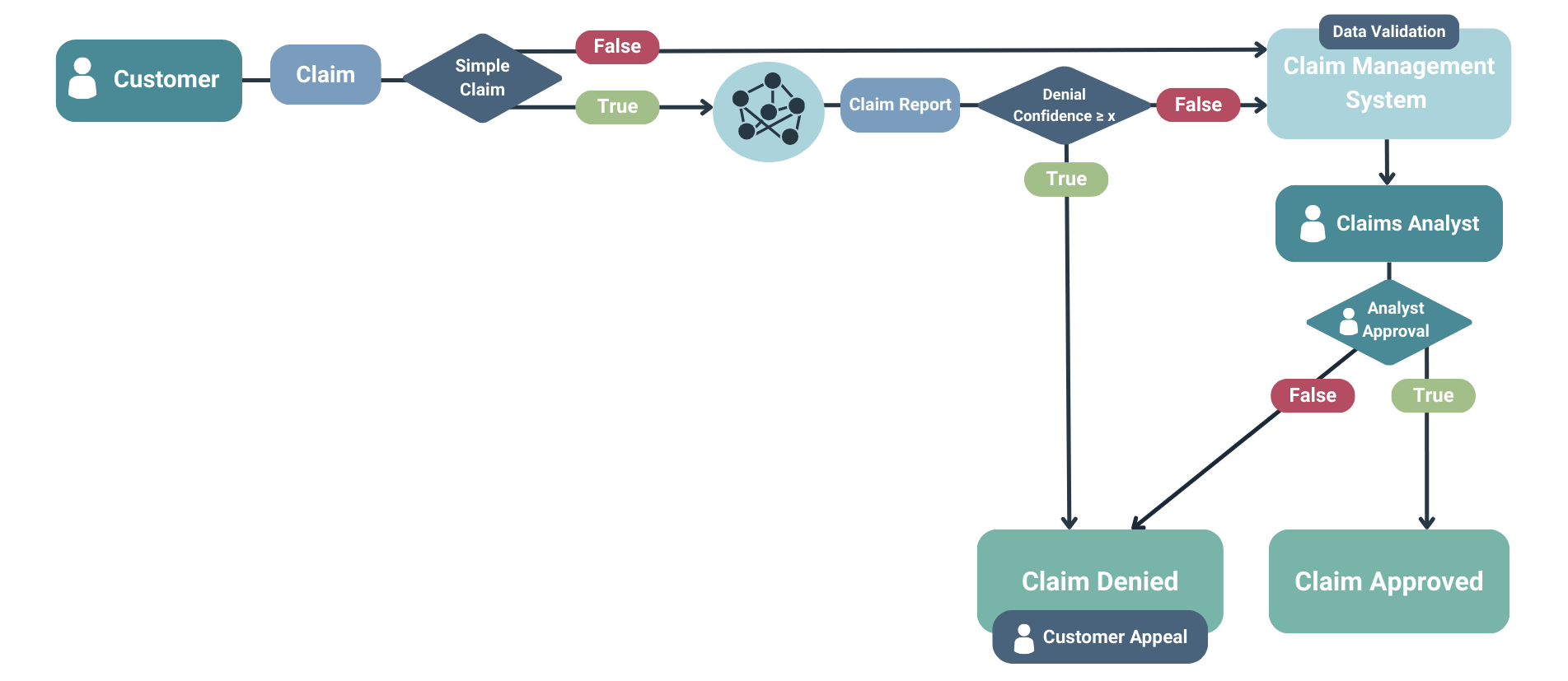}{fig:insurance}
\textbf{Task:} Automotive insurance claims are summarized and reviewed by an AI system.\\
\textbf{Intent:} The AI system increases productivity and profit by reducing the number of simple claims that analysts must review.\\
\textbf{Risks:} There is a risk of incorrectly denying legitimate claims if the model is overconfident and mistaken.\\
\textbf{Human Oversight Level:} Conditionally Autonomous AI\\
\textbf{Institutional Oversight Examples:} Validating data provided (organization best practice), denial confidence threshold choice (organization policy), customer appeal process (organization policy), consumer protection practices (regulation)\\
\textbf{Explanation of AI Workflow:}\\
\textit{Input}: A customer submits an insurance claim. Complex claims go directly to the claims management system; simple claims are sent to the AI system.\\
\textit{Process}: If unconfident, the claim goes to the claims management system for validation. If confident (fits known patterns), the AI system produces a claim report; low-confidence denials are routed for review.\\
\textit{Output}: High-confidence denials are automatically denied; otherwise, a claims analyst reviews and decides on the case.\\
\end{usecase}
\newpage

\noindent\begin{usecase}{7}{Credit Lending Classifier}{Private}{Finance}{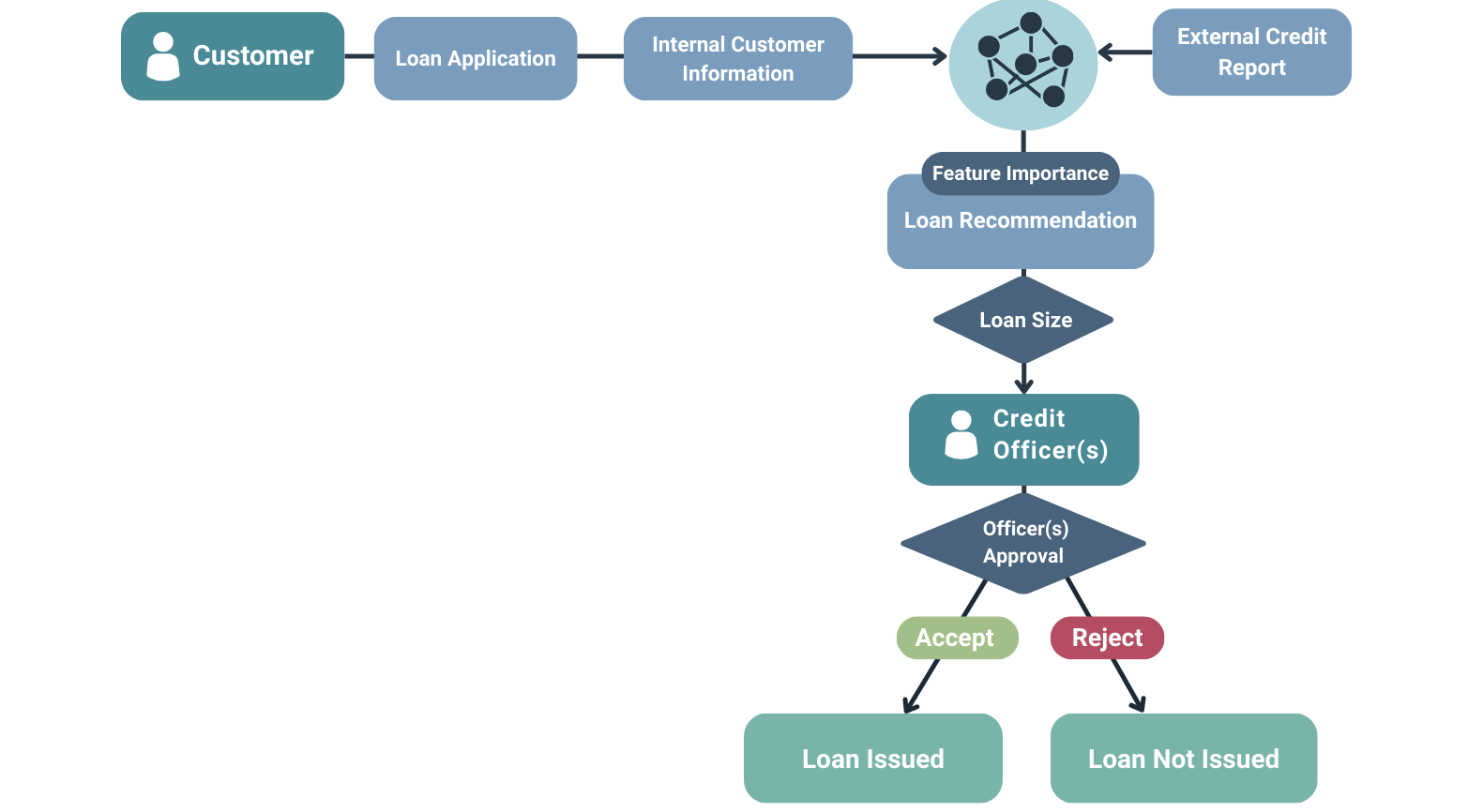}{fig:lending}
\textbf{Task:} The AI system processes loan applications and generates approval recommendations; one or more credit officers review depending on loan size.\\
\textbf{Intent:} The aims of the AI system are to support credit officers and increase efficiency in lending.\\
\textbf{Risks:} The AI system poses potential fairness risks if historical bias is embedded; over-reliance on the system could reduce critical evaluation of borderline applications.\\
\textbf{Human Oversight Level:} Human-Approved AI\\
\textbf{Institutional Oversight Examples:} Number of analysts reviewing the loan (organization policy), lending rules (regulation), data protection (regulation), consumer protections (regulation)\\
\textbf{Explanation of AI Workflow:}\\
\textit{Input}: Loan application data, internal customer information, and external credit reports are fed into the AI system.\\
\textit{Process}: The AI system generates an approval recommendation and feature-importance scores showing how inputs influenced the output.\\
\textit{Output}: One or more officers review the recommendation and make the final decision.\\
\end{usecase}
\newpage

\noindent\begin{usecase}{8}{Physio-Risk Reporter}{Private}{Health}{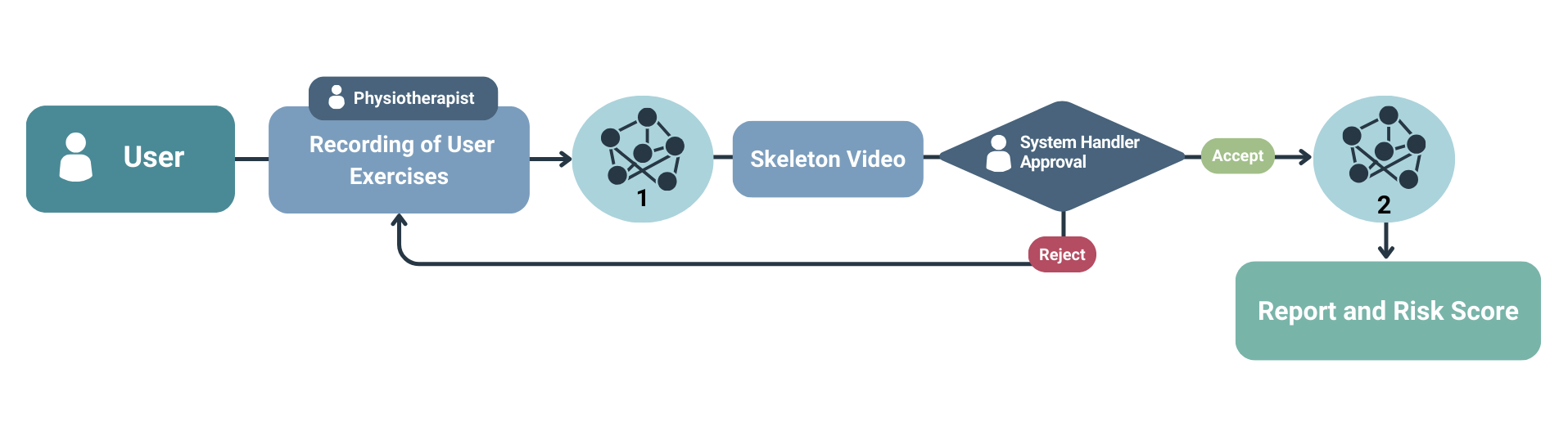}{fig:physio-risk}
\textbf{Task:} An older adult with limited mobility uses the AI system to obtain a physio-related risk assessment; the report is for the user but can be shared with carers. Users may use the AI system multiple times over time.\\
\textbf{Intent:} The AI system provides a private risk score to inform physio-needs, and enables periodic check-ins for users.\\
\textbf{Risks:} The person could injure themselves or become fatigued during exercises.\\
\textbf{Human Oversight Level:} Autonomous AI\\
\textbf{Institutional Oversight Examples:} Ability to redo an exercise up to 3 times (organization best practice), human physiotherapist present (organization policy), data protection (regulation)\\
\textbf{Explanation of AI Workflow:}\\
\textit{Input}: A user completes 12 exercises while being video recorded under supervision; the video goes to an already-trained Model 1.\\
\textit{Process}: Model 1 outputs a skeleton video; a handler checks usability. If approved, a rule-based Model 2 takes the skeleton video as input.\\
\textit{Output}: Model 2 outputs a PDF report for the user with their risk score.\\
\end{usecase}
\newpage

\noindent\begin{usecase}{9}{Writing Assistant}{Private}{Cross-Domain}{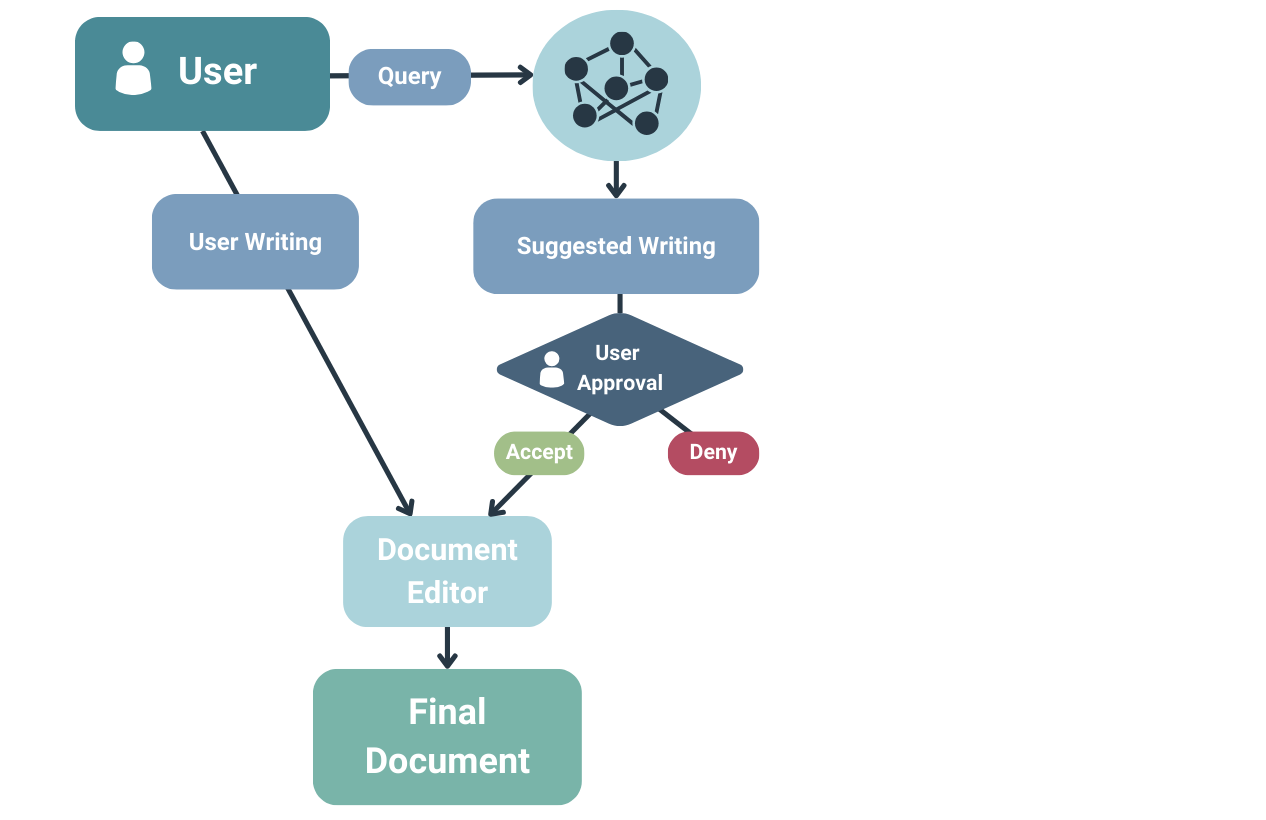}{fig:writing-tool}
\textbf{Task:} Users receive support in writing, brainstorming, and editing their work via an AI system.\\
\textbf{Intent:} The AI system helps users communicate better and support creativity, efficiency, and productivity.\\
\textbf{Risks:} Some of the risks include user over-reliance on the AI system, and the AI system hallucinating or misunderstanding user requests.\\
\textbf{Human Oversight Level:} Human-Led with AI-Assistance\\
\textbf{Institutional Oversight Examples:} Sensitivity filters from the AI provider (ad-hoc practice)\\
\textbf{Explanation of AI Workflow:}\\
\textit{Input}: A user submits a writing or editing request.\\
\textit{Process}: The AI system suggests writing in the editor; the user can accept or deny changes and continue writing.\\
\textit{Output}: The workflow ends when the user is satisfied with the document.\\
\end{usecase}
\newpage

\noindent\begin{usecase}{10}{Hyperparameter Optimizer for LLM and RAG Systems}{Private}{Cross-Domain}{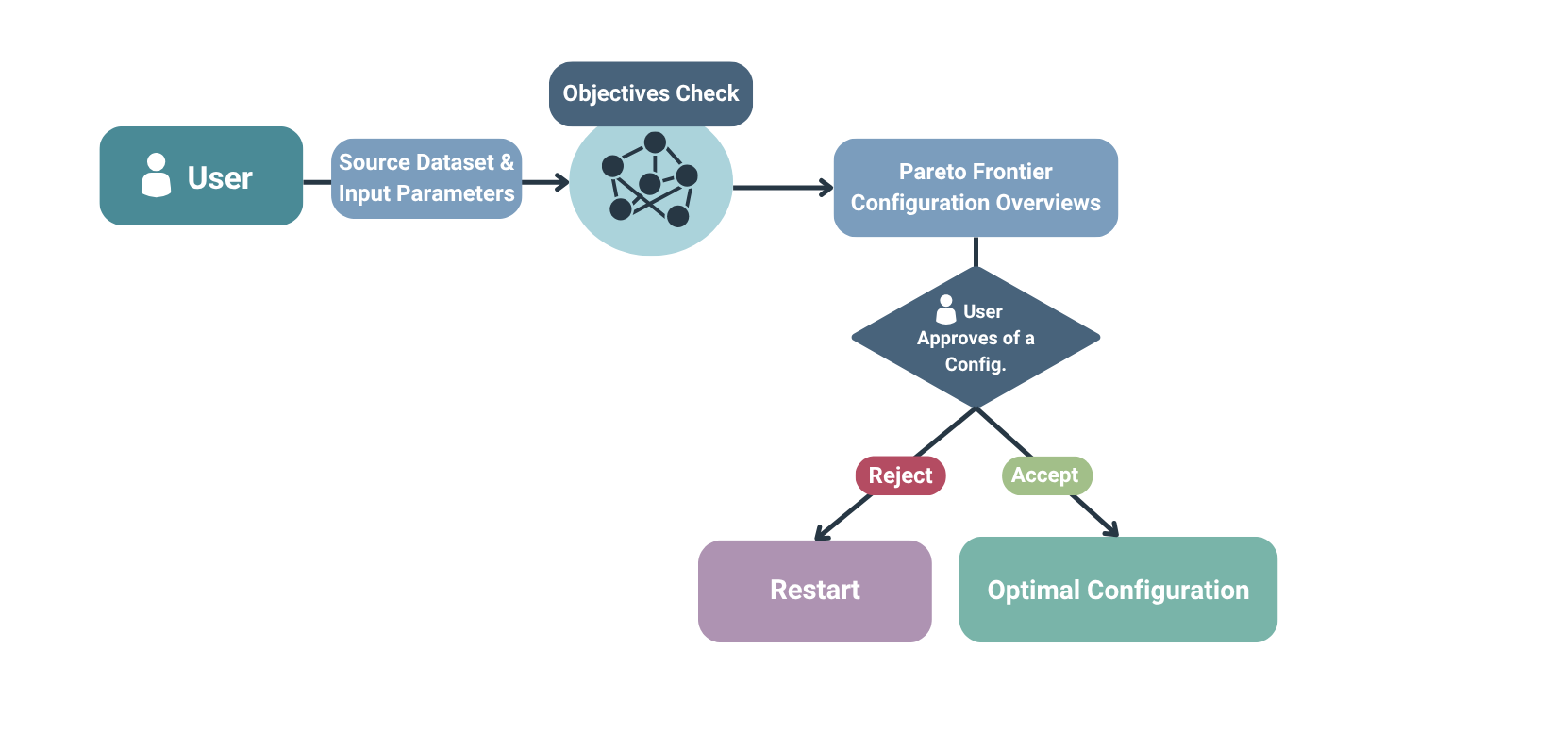}{fig:parameter-optimizer}
\textbf{Task:} The AI system determines optimal configurations for retrieval-augmented generation (RAG) pipelines through multi-objective optimization across safety (hallucinations), alignment (helpfulness), cost, carbon, and latency.\\
\textbf{Intent:} The aims of the AI system are to enable efficient, safer RAG deployments, improve helpfulness, provide benchmarks for compliance, and support informed trade-offs.\\
\textbf{Risks:} Some of the AI system risks include a potential over-reliance on metrics, and that test datasets may not reflect real usage, leading to suboptimal choices.\\
\textbf{Human Oversight Level:} Human-Approved AI\\
\textbf{Institutional Oversight Examples:} Transparent documentation of optimization methods (organization best practice), regular benchmarking (organization best practice), users retain decision authority (organization best practice), safeguards to prevent unsafe configurations (organization best practice)\\
\textbf{Explanation of AI Workflow:}\\
\textit{Input}: Document data, model options, and example queries are given to the AI system.\\
\textit{Process}: The AI system optimizer evaluates configurations on five metrics and visualizes trade-offs.\\
\textit{Output}: A user selects a configuration that balances priorities and can iterate on the configurations by modifying inputs.\\
\end{usecase}
\newpage

\noindent\begin{usecase}{11}{Smart Clinical Triage Tool}{Public}{Health}{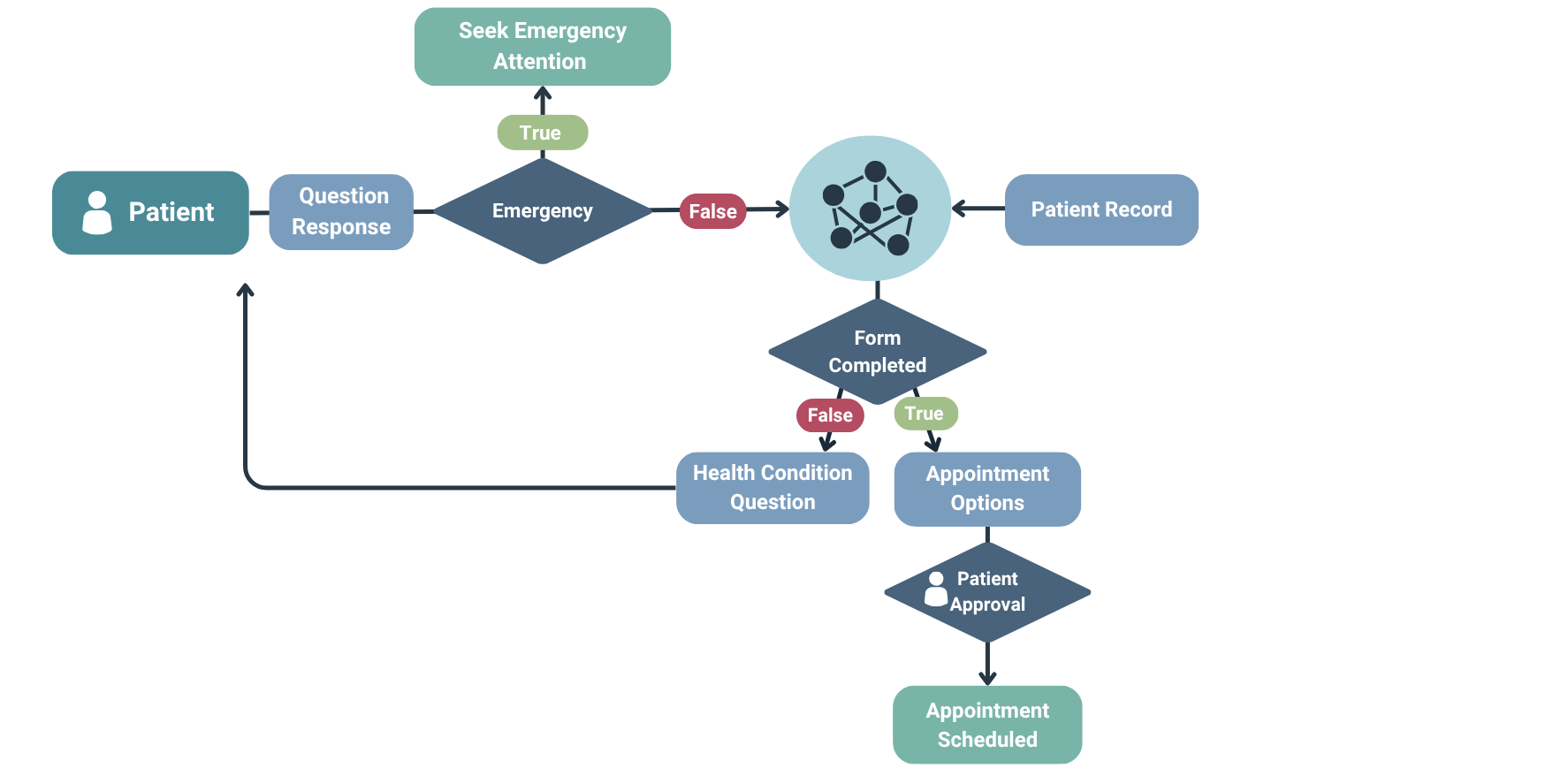}{fig:appt-triage}
\textbf{Task:} The AI system automates clinical triage for general practice, determines urgency, collects structured information, and enables direct booking with appropriate provider.\\
\textbf{Intent:} The goals of the AI system are to reduce administrative burden, enable 24/7 digital triage, and standardize navigation for consistent, equitable outcomes.\\
\textbf{Risks:} A patient could receive an inappropriate classification due to an incomplete or misleading input.\\
\textbf{Human Oversight Level:} Human-Approved AI\\
\textbf{Institutional Oversight Examples:} Audit functionality (ad-hoc practice), mandatory safety netting and escalation in every path (organization best practice), periodic logic review by clinical governance (organization policy), data security (industry standard), procurement assessment (industry standard), digital clinical safety (industry standard), data protection (regulation), medical device compliance (regulation)\\
\textbf{Explanation of AI Workflow:}\\
\textit{Input}: A patient completes an online form via practice website; the AI system may draw on existing clinical records.\\
\textit{Process}: First, an emergency check is done; users with non-emergency cases get adaptive form follow-ups via structured rules from the AI system. The triage concludes with a classification and recommended next step(s).\\
\textit{Output}: The AI system offers appropriate appointment slots and enables booking or messaging without reception staff.\\
\end{usecase}
\newpage

\noindent\begin{usecase}{12}{Chest X-Ray Abnormality Detector}{Public}{Health}{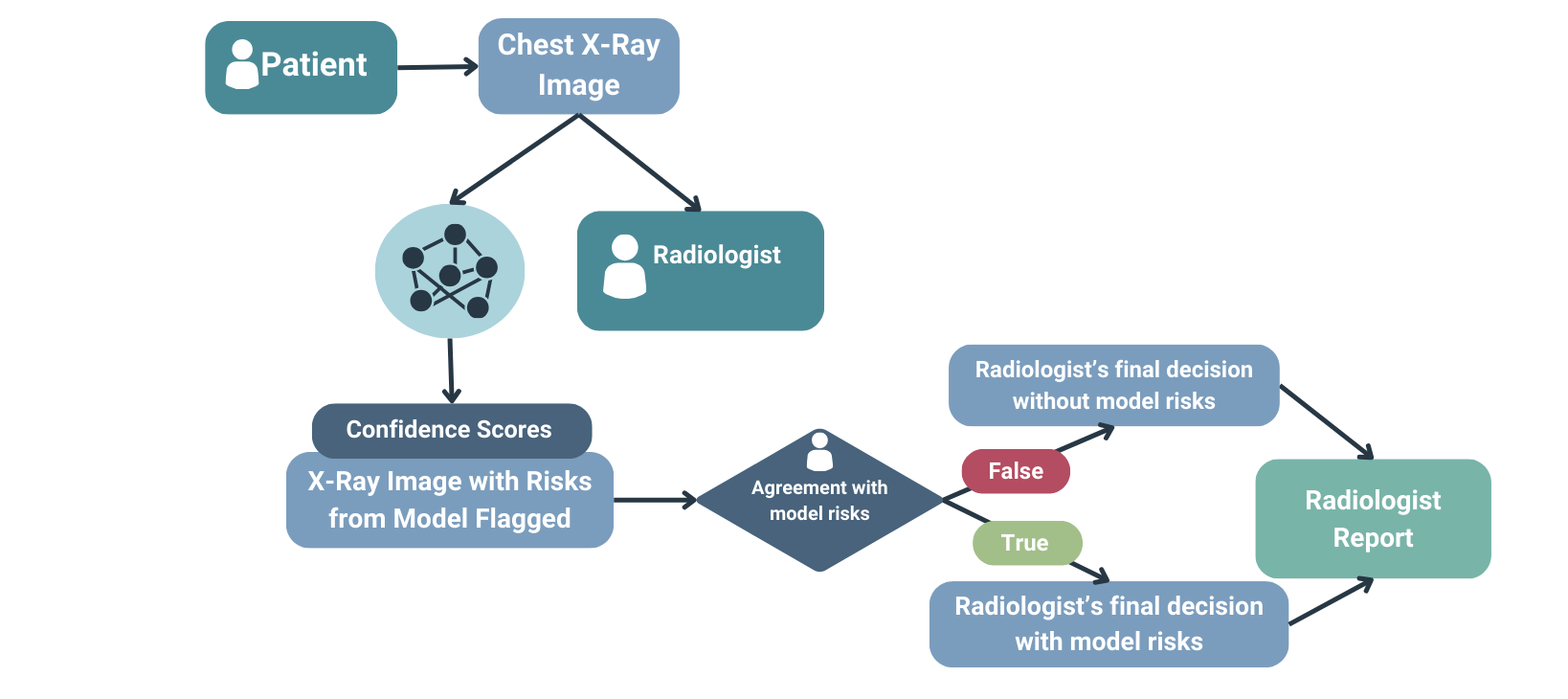}{fig:chest-detector}
\textbf{Task:} The AI system reviews chest X-rays, and flags risks and locations for the radiologist.\\
\textbf{Intent:} The aims of the AI system are to increase speed and accuracy of screening and to prioritize highest-risk patients.\\
\textbf{Risks:} There are some risks with the AI system: automation bias, and model errors, potentially leading to misdiagnosis or missed risks.\\
\textbf{Human Oversight Level:} Human-Led with AI-Assistance\\
\textbf{Institutional Oversight Examples:} Prioritization score to speed review (organization best practice), post-deployment monitoring for automation bias (organization best practice), data protection (regulation), radiation rules (regulation)\\
\textbf{Explanation of AI Workflow:}\\
\textit{Input}: The chest X-ray is sent to the AI system and to the radiologist.\\
\textit{Process}: The AI system overlays flagged regions with confidence scores; radiologist views original first, and then the augmented image.\\
\textit{Output}: The radiologist writes the final report with diagnosis and next steps.\\
\end{usecase}
\newpage 

\noindent\begin{usecase}{13}{Mental Health Triage Tool}{Public}{Health}{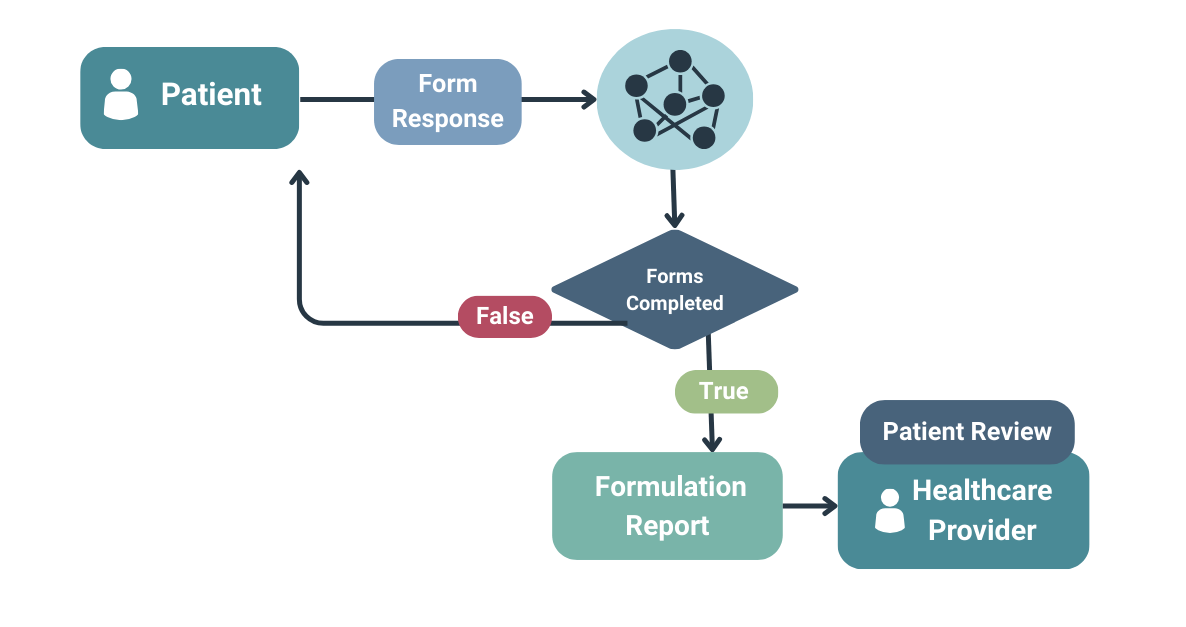}{fig:mental-health}
\textbf{Task:} The AI system generates a preliminary patient report for providers based on responses to intake forms.\\
\textbf{Intent:} The aims of the AI system are to improve intake efficiency and thoroughness.\\
\textbf{Risks:} Patients could be mislabeled if they answer incorrectly or skip questions.\\
\textbf{Human Oversight Level:} Autonomous AI\\
\textbf{Institutional Oversight Examples:} Lived-experience digital group sign-off (organization policy), project and clinical boards sign-off (organization policy), digital clinical safety standards and assurance (industry standard), equality and health impact assessment (regulation), data protection (regulation), medical device compliance (regulation)\\
\textbf{Explanation of AI Workflow:}\\
\textit{Input}: A patient completes mental health intake forms.\\
\textit{Process}: The AI system continues providing forms until completed, then generates a formulation report.\\
\textit{Output}: The report is reviewed by the healthcare provider with the patient.\\
\end{usecase}
\newpage

\noindent\begin{usecase}{14}{CT Scan Risk Detector}{Public}{Health}{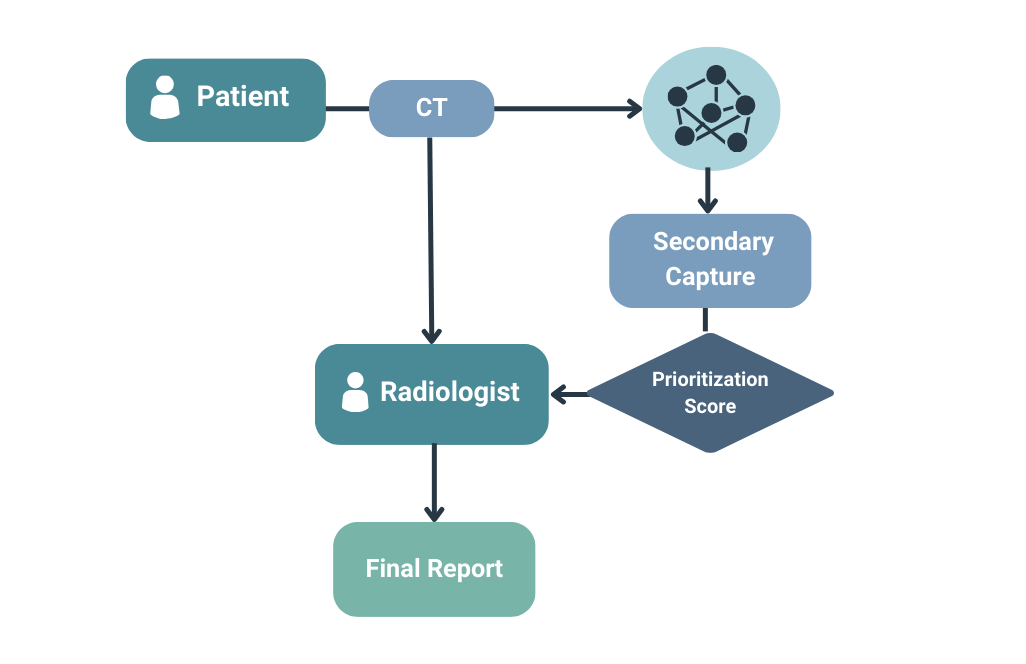}{fig:ct-scan}
\textbf{Task:} For a patient CT scan, the AI generates a secondary view with flagged concerns and a prioritization score to speed review by a radiologist.\\
\textbf{Intent:} The AI system supports radiologists in detecting risks and prioritizing urgent cases.\\
\textbf{Risks:} Radiologists could become over reliant on the system’s secondary capture if they do not view the initial scans first. If the system is incorrect, it could fail to prioritize urgent cases, which is how scans were viewed before this system was implemented. \\
\textbf{Human Oversight Level:} Human-Led with AI-Assistance\\
\textbf{Institutional Oversight Examples:} Prioritization score (organization best practice), data opt-out scheme (organization policy), data protection (regulation)\\
\textbf{Explanation of AI Workflow:}\\
\textit{Input}: A CT scan is sent to the AI system and to the radiologist.\\
\textit{Process}: The AI system generates a secondary capture image and prioritization score.\\
\textit{Output}: The radiologist reviews both images and writes the final report.\\
\end{usecase}
\newpage 

\noindent\begin{usecase}{15}{Consultation AI Note-Taker}{Public}{Health}{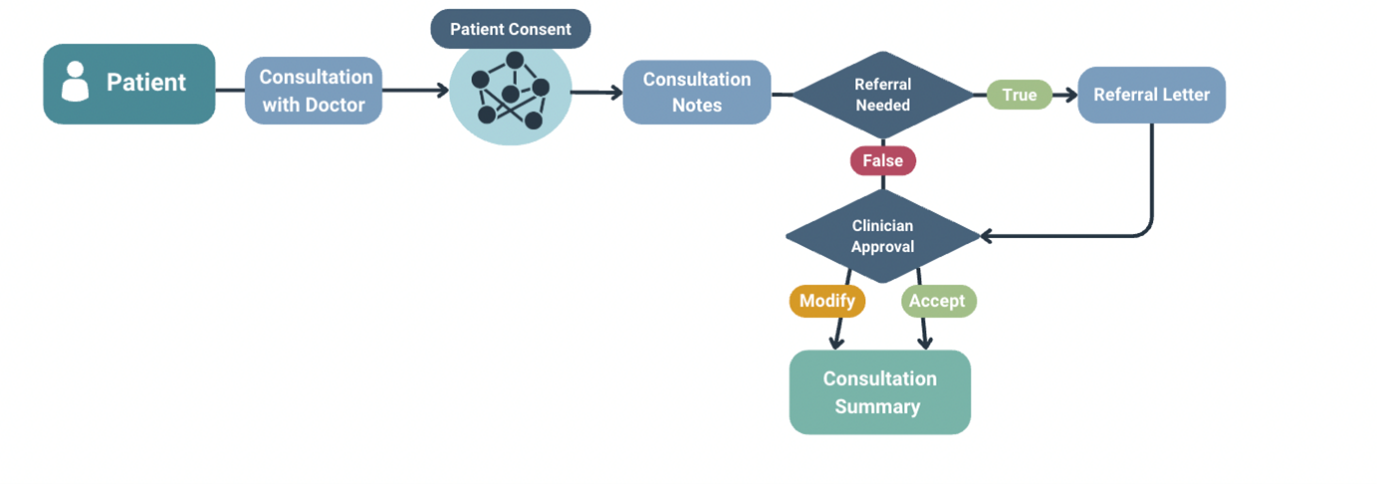}{fig:note-taker}
\textbf{Task:} During a patient–doctor consultation, the system generates notes and a referral letter (if needed) for the doctor to review and use.\\
\textbf{Intent:} The AI system improves note-taking efficiency and thoroughness.\\
\textbf{Risks:} Clinicians could become over-reliant on the system for documentation.\\
\textbf{Human Oversight Level:} Human-Led with AI-Assistance\\
\textbf{Institutional Oversight Examples:} Digital clinical safety practices (industry standard), data protection (regulation), equality and health impact assessment (regulation)\\
\textbf{Explanation of AI Workflow:}\\
\textit{Input}: With patient approval, the patient-doctor consultation is recorded and passed to the AI system.\\
\textit{Process}: The AI system generates consultation notes and predicts if a referral is needed; if so, the system drafts a letter.\\
\textit{Output}: Clinician reviews/edits notes and adds them to the chart; if a referral is appropriate, clinician reviews/edits the letter before sending.\\
\end{usecase}
\newpage

\noindent\begin{usecase}{16}{Carer-AI Kit Risk Assessor}{Public}{Health}{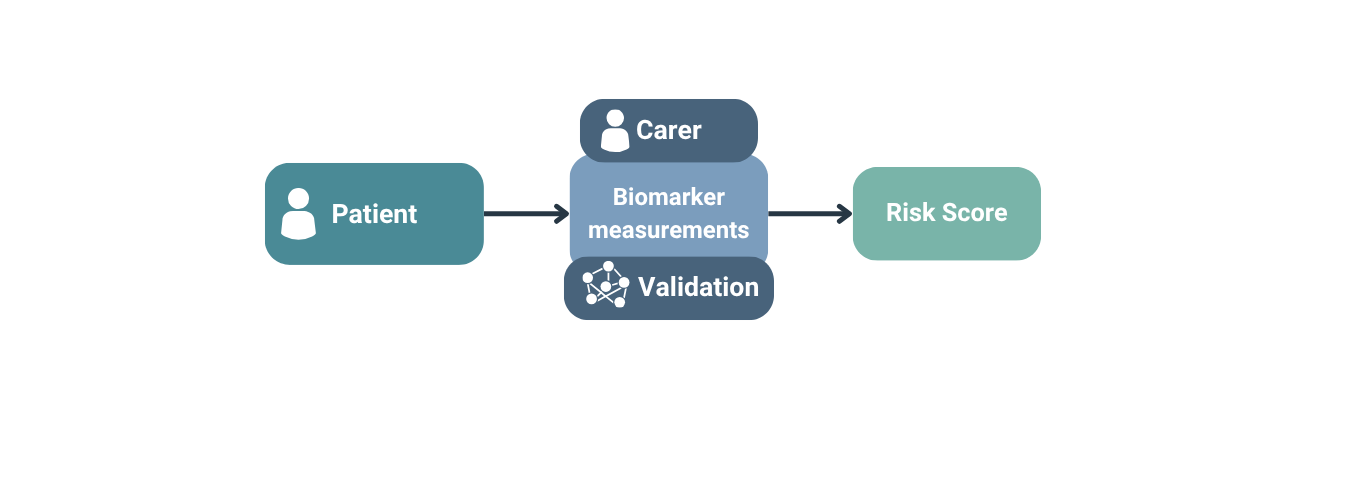}{fig:care-ai}
\textbf{Task:} The AI system is apart of a platform used to validate a carer’s biomarker measurements (e.g., heart rate, pulse, blood pressure) and to provide a deterioration risk score.\\
\textbf{Intent:} The AI system supports accurate biomarker measurement by flagging retakes and provides a risk score that helps determine next actions.\\
\textbf{Risks:} Use of this AI system could lead to carers lacking confidence in readings or device usage. Other AI system risks include medical device bias and data safety risks.\\
\textbf{Human Oversight Level:} Autonomous AI\\
\textbf{Institutional Oversight Examples:} Digital clinical safety practices (industry standard), data protection (regulation), medical device compliance (regulation), healthcare compliance (regulation)\\
\textbf{Explanation of AI Workflow:}\\
\textit{Input}: A carer takes biomarker measurements.\\
\textit{Process}: The AI system validates measurement quality and flags retakes if needed.\\
\textit{Output}: The AI system generates a risk score for the carer; if needed, the carer takes further action (e.g., escalate for medical help).\\
\end{usecase}
\newpage 

\noindent\begin{usecase}{17}{OriginTrail Decentralized Knowledge Graph (DKG)}{Private}{Cross-Domain}{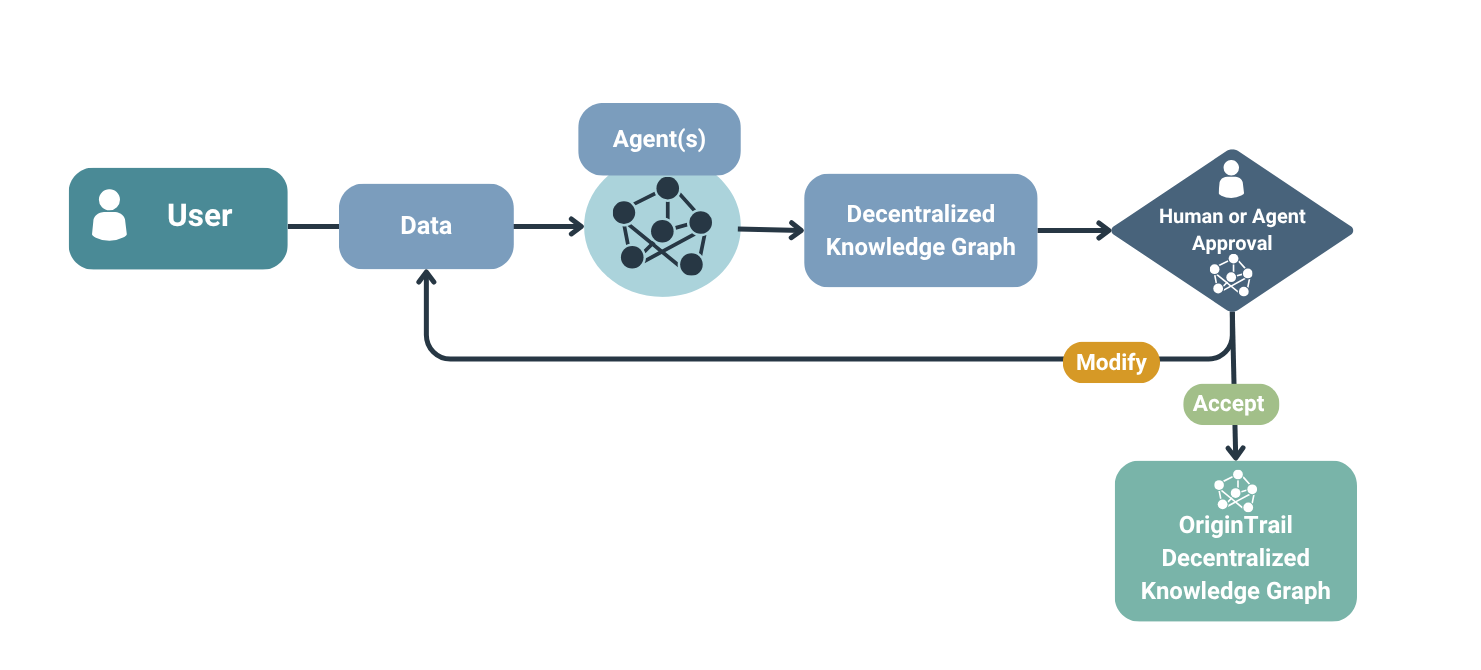}{fig:origintrail}
\textbf{Task:} OriginTrail DKG enables people and agents to turn data into verifiable, structured, interconnected knowledge that remains traceable to its source.\\
\textbf{Intent:} The AI system combines symbolic AI (a DKG) with GenAI to create a resilient decentralized AI infrastructure with source traceability.\\
\textbf{Risks:} It is possible for agents to make a mistake while building the graph; an approval agent or human may miss an error and become over-reliant.\\
\textbf{Human Oversight Level:} Conditionally Autonomous AI\\
\textbf{Institutional Oversight Examples:} User/community-led verification of the knowledge graph (ad-hoc practice), open-source code (organization policy)\\
\textbf{Explanation of AI Workflow:}\\
\textit{Input}: A user submits data (e.g., PDFs) to the AI system.\\
\textit{Process}: Agents transform data into a knowledge graph draft (nodes and connections) for the DKG; roles may include coordinator or ontology expert.\\
\textit{Output}: A review gateway (a human or an agent) can approve/adjust the DKG before publishing; once published, the DKG gains cryptographic source verifiability and links with other knowledge on the network.\\
\end{usecase}
\newpage

\noindent\begin{usecase}{18}{Automated Imaging Protocol Selector}{Public}{Health}{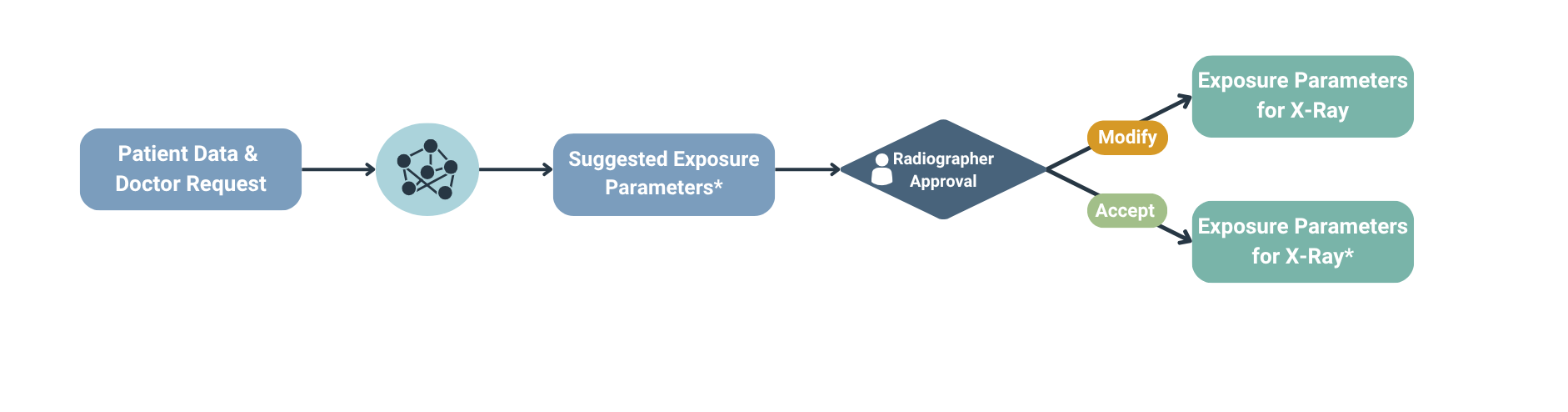}{fig:auto}
\textbf{Task:} The AI system automatically selects radiology exposure parameters based on patient information to support the radiographer.\\
\textbf{Intent:} The AI system selects parameters that enable diagnostic reading while limiting exposure to ionizing radiation. This parameter selection is crucial because if too-low of a dose is given, that could result in retakes and false negatives; on the other hand, if too-high of a dose is given, that would likely result in harm.\\
\textbf{Risks:} In extreme cases, the AI system could lead to excessive ionizing radiation which can cause cell damage and potentially cancer or, at extreme levels, death.\\
\textbf{Human Oversight Level:} Human-Led with AI-Assistance\\
\textbf{Institutional Oversight Examples:} Standard operating procedures on equipment settings and staff competency (organization best practice), automatic exposure control and acceptable dose levels (industry best practice), patient communication and documentation (industry standard), radiation protection standards (regulation), personnel qualification (regulation), equipment testing/quality control/monitoring (regulation), informed consent (regulation)\\
\textbf{Explanation of AI Workflow:}\\
\textit{Input}: The AI system is given patient data and the doctor’s intake information.\\
\textit{Process}: The rule-based AI system uses radiographer input and connected systems (e.g., electronic patient record and radiology information system) with manufacturer thresholds, possibly adjusted by installers, medical physicists, or quality control staff per guidance/law.\\
\textit{Output}: A radiographer reviews the AI system output and may modify suggested settings before acquiring the image.\\
\end{usecase}
\newpage

\noindent\begin{usecase}{19}{Call Center Virtual Assistant}{Public}{Public Services}{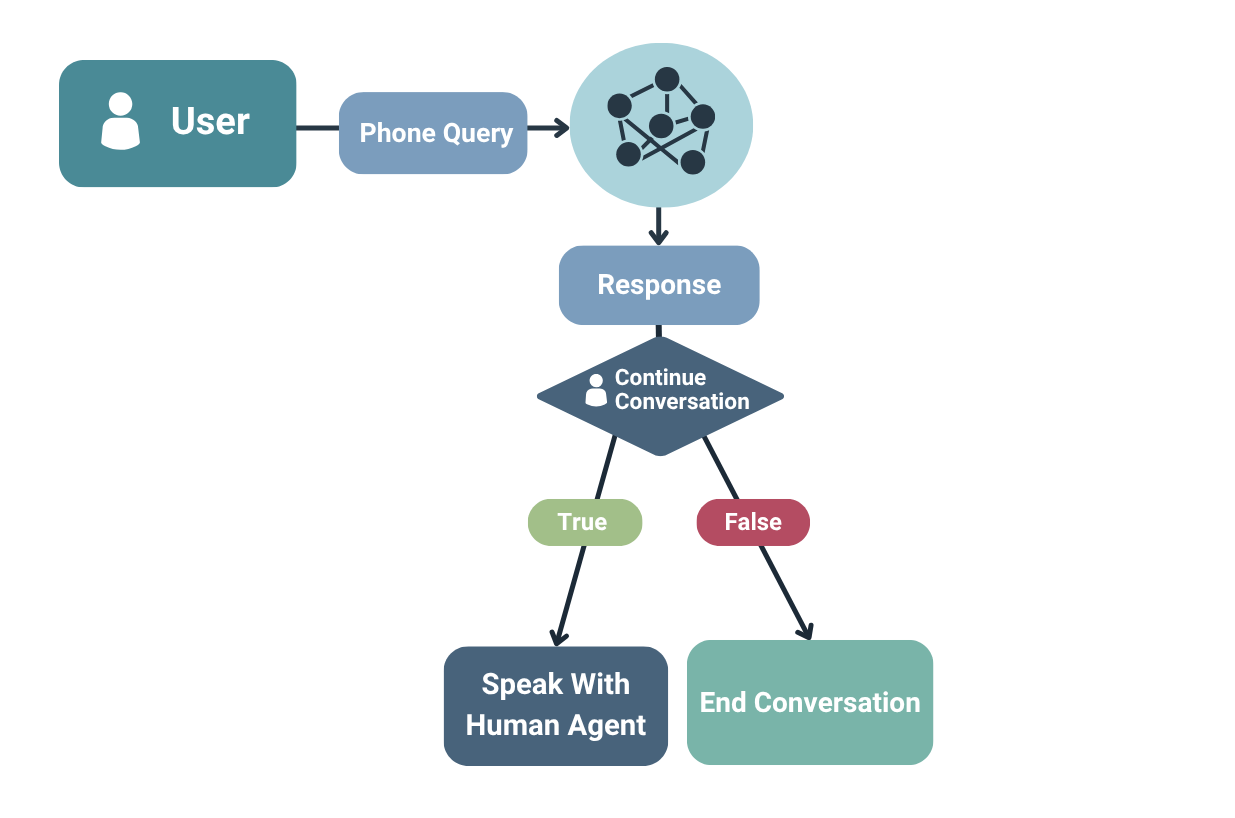}{fig:call-center}
\textbf{Task:} The AI system is a virtual assistant that processes service requests over the phone (e.g., payment systems) and answers user-specific inquiries 24/7.\\
\textbf{Intent:} The goals of the AI system are to reduce administrative workload and provide reliable support for users during and after business hours without increasing costs.\\
\textbf{Risks:} The AI system could misinform users by generating incorrect or incomplete responses; the system may fail to understand the user’s request.\\
\textbf{Human Oversight Level:} Conditionally Autonomous AI\\
\textbf{Institutional Oversight Examples:} Privacy by design (organization best practice), fine-tuning (organization best practice), post-deployment monitoring (organization best practice), ethics impact assessment (organization policy), inclusive multimodal access mandate (organization policy), AI procurement practices (organization policy), data security (industry standard), risk assessment and management (regulation), data protection (regulation), automated decision-making transparency (regulation)\\
\textbf{Explanation of AI Workflow:}\\
\textit{Input}: A user calls the support line and asks a question.\\
\textit{Process}: The AI system uses speech-to-text, natural language understanding, and retrieval to generate responses.\\
\textit{Output}: If the AI system understands and retrieves the necessary details (e.g., the status of a payment), it replies via text-to-speech; otherwise, it hands off to a human during business hours, or asks the user to call back in-hours.\\
\end{usecase}
\newpage

\noindent\begin{usecase}{20}{Image Blurring Tool}{Public}{Public Services}{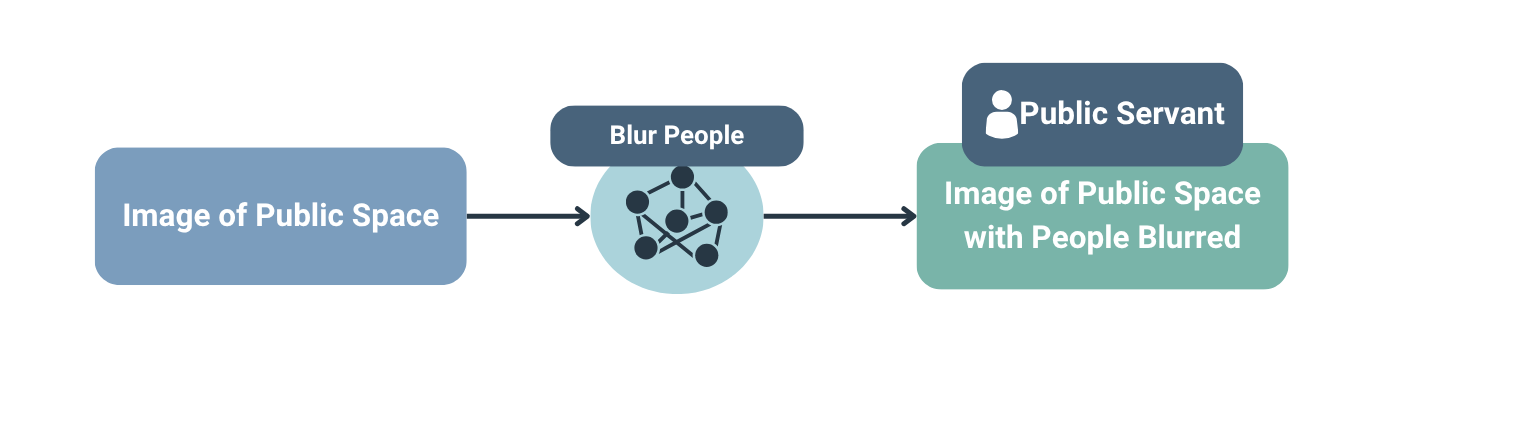}{fig:blurring} 
\textbf{Task:} The AI system blurs images of people in public areas to protect privacy; blurred images are then used for downstream public-service tasks.\\
\textbf{Intent:} The aims of the AI system are to maintain the privacy of individuals captured on cameras used to support public services and to enable multiple use cases on top of blurred imagery.\\
\textbf{Risks:} Some risks of the AI system include potential for bias against certain groups; false positives (blur non-people) or false negatives (miss people).\\
\textbf{Human Oversight Level:} Autonomous AI\\
\textbf{Institutional Oversight Examples:} Ethics impact assessment (organization policy), bias assessment (organization policy), internal risk assessment (regulation), data protection (regulation), algorithm register for transparency (regulation)\\
\textbf{Explanation of AI Workflow:}\\
\textit{Input}: A vehicle captures photos of public areas; originals with people are not saved once sent to the AI system.\\
\textit{Process}: Computer vision detects people and applies a blurring overlay.\\
\textit{Output}: The blurred image is used as a privacy-preserving photo of a public space for other tasks; missed blurs can be flagged and the system retrained later.\\
\end{usecase}
\newpage 

\section{Oversight Levels Questionnaire}
\label{appendix:questions}

To detect what level of human oversight an AI system abides by, we recommend reviewing Table \ref{tab:oversight-questions}. 
The level of human oversight, risk (e.g., EU AI Act risk categories), and domain all factor into what institutional oversight should be implemented before deployment. 
When we consider how the use case domain and risk category factor into institutional oversight, the below questions are meant to help an individual or team brainstorm what oversight is potentially needed. We take inspiration from the EU AI Act for the risk categories. They are a starting point for practitioners who could discuss with their teams. We note that top-down guidance on risk tolerance and how governance is implemented to respond to risks is likely key. 
\begin{itemize}
    \item Is the AI system high-risk such that it can have severe implications on people's access to essential services (e.g., safety, fundamental rights, or health)? If yes, then the practitioner should consider implementing the strictest levels of institutional oversight. 
    \item Is the the AI system a limited risk (e.g., an internally facing chatbot)? If so, unless specific regulations must be followed, the institutional oversight needed at the most is likely medium levels (e.g., organization best practice, organization policy, industry standards). 
    \item Is the AI system minimal or no-risk? If so, the strictest forms of institutional oversight are likely not required. At the least, low to medium forms of institutional oversight could be implemented like ad-hoc practices and organization best practices. 
\end{itemize}

\vspace{-0.6em}
\begin{table*}[!ht]
\centering
\renewcommand{\arraystretch}{0.8}
\resizebox{\textwidth}{!}{%
\begin{tabular}{|p{0.3\textwidth}|*{4}{>{\centering\arraybackslash}p{0.17\textwidth}|}}
\hline
\textbf{Question} & \textbf{Autonomous AI} & \textbf{Conditionally Autonomous AI} & \textbf{Human-Approved AI} & \textbf{Human-Led with AI-Assistance} \\
\hline
Is the AI output the final output? & \textbf{\checkmark} & \textbf{\checkmark} & \textbf{\checkmark} & \\
\hline
Is there a pathway where the system can act autonomously without any human oversight? & \textbf{\checkmark} & \textbf{\checkmark} & & \\
\hline
Does a pathway exist where a human has to check the AI output or decides on the final output? & & \textbf{\checkmark} & \textbf{\checkmark} & \textbf{\checkmark} \\
\hline
Can a human accept the AI output as the final output or reject it? & & & \textbf{\checkmark} & \textbf{\checkmark} \\
\hline
Does a human have the chance to review, use, or modify the AI output in a final output? & & & & \textbf{\checkmark} \\
\hline
\end{tabular}%
}

\caption{These are questions to support practitioners in distinguishing under which human oversight level their AI use case falls. If your AI use case checks all of the boxes under a certain column, then that is the correct human oversight level.}
\label{tab:oversight-questions}
\end{table*}